\documentclass[journal]{IEEEtran}
\usepackage{xurl}
\usepackage{hhline}

\usepackage{multirow}
\usepackage{ragged2e}
\usepackage{acronym}
\usepackage{indentfirst}

\usepackage{comment}
\usepackage{makecell}
\usepackage{relsize}
\usepackage{hhline}

\usepackage[table]{xcolor} 
\usepackage{multirow} 
\usepackage{colortbl} 
\usepackage{hhline} 
\usepackage{array} 
\usepackage[colorinlistoftodos,prependcaption,textsize=tiny]{todonotes}
\usepackage{soul}
\usepackage{cite}
\usepackage{amsmath,amssymb,amsfonts}
\usepackage{algorithmic}
\usepackage{graphicx}
\usepackage{textcomp}
\usepackage{lipsum}
\usepackage[utf8]{inputenc}
\usepackage{textcomp}
\usepackage{caption}
\usepackage{cite}
\usepackage{authblk}
\usepackage{amsmath}
\usepackage{nomencl}
\usepackage{enumitem}
\usepackage{epsfig}
\usepackage{booktabs}
\usepackage{upgreek}
\usepackage[normalem]{ulem}
\useunder{\uline}{\ul}{}
\usepackage{longtable}
\usepackage{array}
\usepackage{tabularx}
\usepackage{xcolor}

\usepackage{rotating,multirow,multicol}
\usepackage{wasysym}
\usepackage{upgreek}
\usepackage[flushleft]{threeparttable}
\usepackage[T1]{fontenc}
\usepackage{newtxmath}
\usepackage{euscript}
\DeclareMathAlphabet{\mathpzc}{T1}{pzc}{m}{it}
\usepackage{pifont}
\usepackage{gensymb}

\usepackage{colortbl}
\definecolor{lightGreen}{RGB}{169, 209, 142}
\definecolor{morelightGreen}{RGB}{226, 240, 217}
\usepackage{fancyhdr}
\definecolor{myGreen}{rgb}{0.8,1,0.8} 

\def\BibTeX{{\rm B\kern-.05em{\sc i\kern-.025em b}\kern-.08em
    T\kern-.1667em\lower.7ex\hbox{E}\kern-.125emX}}

\usepackage{orcidlink}

\begin{document}

 \title{Multi-Satellite MIMO Systems for Direct User-Satellite Communications: A Survey}

\author{Zohre~Mashayekh~Bakhsh$^\text{\orcidlink{0000-0002-3798-968X}}$,
        Yasaman~Omid$^\text{\orcidlink{0000-0002-5739-8617}}$,~\IEEEmembership{Student Member,~IEEE,}
        Gaojie~Chen$^\text{\orcidlink{0000-0003-2978-0365}}$,~\IEEEmembership{Senior Member,~IEEE,}
        Farbod~Kayhan$^\text{\orcidlink{0000-0002-1733-5435}}$,
        Yi~Ma$^\text{\orcidlink{0000-0002-6715-4309}}$,~\IEEEmembership{Senior Member,~IEEE,}
        and Rahim~Tafazolli$^\text{\orcidlink{0000-0002-6062-8639}}$,~\IEEEmembership{Senior Member,~IEEE}
 \thanks{Zohre Mashayekh Bakhsh, Gaojie Chen, Yi Ma, and Rahim Tafazolli are with the Institute for Communication Systems, Home of
the 6G Innovation Centre, University of Surrey, GU2 7XH Guildford, U.K. 
Yasaman Omid and Farbod Kayhan were with the Institute for Communication Systems, Home of
the 6G Innovation Centre, University of Surrey, GU2 7XH Guildford, U.K. Currently, Yasaman Omid is with Loughborough University, and Farbod Kayhan is the Lead SatCom 5G R\&D Engineer at Telespazio, UK.
(E-mails: z.mashayekhbakhsh@surrey.ac.uk, y.omid@lboro.ac.uk, gaojie.chen@ieee.org, farbod.kayhan@telespazio.com, y.ma@surrey.ac.uk, r.tafazolli@surrey.ac.uk) (\it Corresponding author: Gaojie Chen.)}%

}

\markboth{IEEE COMMUNICATIONS SURVEYS $\And$ TUTORIALS,~Vol.~X, No.~X, XXX}%
{Shell \MakeLowercase{\textit{et al.}}: Bare Demo of IEEEtran.cls for IEEE Communications Society Journals}

\maketitle

\begin{abstract}
\textcolor{black}{
Advancements in satellite technology have made direct-to-device connectivity a viable solution for ensuring global access. This method is designed to provide internet connectivity to remote, rural, or underserved areas where traditional cellular or broadband networks are lacking or insufficient.}
This paper is a survey providing an in-depth review of multi-satellite Multiple Input Multiple Output (MIMO) systems as a potential solution for addressing the link budget challenge in direct user-satellite communication.
Special attention is given to works considering multi-satellite MIMO systems, both with and without satellite collaboration. In this context, collaboration refers to sharing data between satellites to improve the performance of the system.
\textcolor{black}{
This survey begins by explaining several fundamental aspects of satellite communications (SatComs), which are vital prerequisites before investigating the multi-satellite MIMO systems. These aspects encompass satellite orbits, the structure of satellite systems, SatCom links, including the inter-satellite links (ISL) which facilitate satellite cooperation, satellite frequency bands, satellite antenna design, and satellite channel models, which should be known or estimated for effective data transmission to and from multiple satellites.
Furthermore, this survey distinguishes itself by providing more comprehensive insights in comparison to other surveys. It specifically delves into the Orthogonal Time Frequency Space (OTFS) within the channel model section. It goes into detail about ISL noise and channel models, and it extends the ISL section by thoroughly investigating hybrid FSO/RF ISLs.} Furthermore, analytical comparisons of simulation results from these works are presented to highlight the advantages of employing multi-satellite MIMO systems.
\end{abstract}


\printnomenclature[2cm]

\IEEEpeerreviewmaketitle

\section{Introduction}
\label{introduction}
\IEEEPARstart{W}{hile} terrestrial mobile networks have been widely deployed, there are still economically challenging areas around the world that lack adequate connectivity. Providing coverage in rural or remote regions, for example, poses financial obstacles as the return on investment may not justify the costs involved. In contrast, Satellite Communication (SatCom) offers a cost-effective solution. For instance Geostationary Earth Orbit (GEO) with a single satellite capable of covering vast geographic areas. This makes SatComs an attractive option for enhancing connectivity in rural and remote regions \cite{lin20215g, saarnisaari20206g}.

The combination of satellite and terrestrial networks known as Space Air Ground Integrated Network (SAGIN), holds great promise for meeting the demanding requirements of Fifth Generation (5G)/Sixth Generation (6G) systems\cite{Reinerwhitepaper,rinaldi2020non}. These requirements include global and seamless connectivity, increased throughput, and the ability to support mission-critical services \cite{dahlman20145g}.
Addressing the initial requirement mentioned, satellites, with their extensive coverage capabilities, can effectively complement and expand the existing congested terrestrial networks, serving both densely populated and rural areas \cite{gaber20205g, guidotti2019architectures}.
As for the second requirement mentioned above, multi-satellite systems have emerged as a potential solution to enhance data throughput. Therefore, in this survey, we aim to concentrate our investigation on multi-satellite systems.

Additionally, satellites play a vital role in ensuring integration, backhaul, resilience, security, availability, Internet of Things (IoT) support, and efficient spectrum utilization in the context of 5G \cite{evans2014role}.
The 3rd Generation Partnership Project (3GPP) has initiated a study item for the 5G Non-Terrestrial Networks (NTNs), aiming to integrate satellite systems into mobile broadband and machine-type communication scenarios \cite{3GPPrel15}. Other studies such as \cite{3GPPTR38913, 3GPPTR38822, 3GPPTR38821, 3GPPTR22822, sirotkin20205g}, highlight 3GPP's ongoing efforts to investigate and advance the integration of SatComs within the 5G framework.
To seamlessly incorporate NTN into the 5G system, certain adaptations are necessary, particularly within the Radio Access Network (RAN). In the RAN, the conventional terrestrial Base Stations (BSs) are back-hauled via SatCom links instead of using fiber optics. 

\textcolor{black}{One of the challenges in SatCom is closing the link budget. The performance of a link is typically assessed by evaluating the ratio of the received carrier power (C) to the noise power spectral density ($N_0$), known as Carrier-to-Noise Ratio (CNR). This ratio serves as an important indicator of the link quality. In the context of digital communications, the CNR value of the participating links plays a crucial role in determining the QoS, often measured by the Bit Error Rate (BER) \cite{maral2020satellite}.
Another significant parameter in the link budget is the bandwidth occupied by the carrier. This bandwidth depends on
various factors, including the data rate, channel coding rate (forward error correction), and the modulation scheme.
Achieving an optimal balance between the required carrier power and the occupied bandwidth is essential for cost-effective link design in SatComs \cite{maral2020satellite}.
In the context of 3GPP, there have been several discussions regarding link budget analysis. These discussions involve studies with specific parameter examples and simulation outcomes, as documented in  3GPP TR 36.763 \cite{3GPPTR36763} and the 3GPP Release 15 \cite{3GPPrel15}.}

Another challenge in the adaptation of SatComs especially GEO SatComs, for 5G is the higher latency compared to terrestrial communication. This increased delay can impact the performance and responsiveness of 5G services, which are designed to support real-time applications and other latency-sensitive tasks. 
To deal with the delay problem, 3GPP suggests introducing new Radio Access Technology that enables core network functions to select the most suitable RAN for specific Quality of Service (QoS) needs \cite{kaloxylos2018survey}. An amendment proposal suggests a recommendation to inform the policy control function and application function regarding the usage of satellite access or backhaul. This is done to account for the possibility of longer terrestrial Round Trip Times (RTT) when utilizing satellite connections. By providing this information, the network can better handle the potential delays associated with SatCom and adjust its operations accordingly. In SatComs, there is a need to prevent connection timeouts that can occur due to delayed response messages. To address this issue, it becomes necessary to increase the duration of the existing timers used for mobility and session management. By extending these timers, the network allows more time for the transmission and reception of messages, accommodating the additional latency associated with SatComs and avoiding premature connection timeouts \cite{Reinerwhitepaper}.

\begin{table*}[ht]
  \centering
  \caption{COMPARISON WITH OTHER RELATED SURVEYS}
\label{tab1}
\begin{tabular}
{|p{5cm}|p{0.535cm}|p{0.535cm}|p{0.535cm}|p{0.535cm}|p{0.535cm}|p{0.535cm}|p{0.535cm}|p{0.535cm}|p{0.535cm}|p{1.5cm}|}
    \hline
    \rowcolor{lightGreen}
    \multicolumn{1}{|c|}{\cellcolor{morelightGreen}\textbf{Covered Topics}} & \multicolumn{10}{c|}{\cellcolor{lightGreen}\textbf{References}} \\
    \hhline{{\arrayrulecolor{black}}|-|-|-|-|-|-|-|-|-|-|-|}
    \rowcolor{lightGreen}
    \cellcolor{morelightGreen} & \cite{al2022survey} 2023 & \cite{centenaro2021survey} 2021 & \cite{arapoglou2010mimo} 2011 & \cite{prol2022position} 2022 & \cite{kodheli2020satellite} 2021 & \cite{abo2019survey} 2019 & \cite{Baeza2022Overview} 2022 & \cite{muri2012survey} 2012 & \cite{heo2023mimo} 2023 & Our Survey \\
    \hline
    \rowcolor{lightGreen}
    \cellcolor{morelightGreen} ISL Comparison & \checkmark & \checkmark & $\times$ & $\times$ & \checkmark & $\times$ & $\times$ & $\times$ & $\times$ & \checkmark \\
    \rowcolor{lightGreen}
    \cellcolor{morelightGreen} ISL Channel Model & $\times$ & $\times$ & $\times$ & $\times$ & $\times$ & $\times$ & $\times$ & $\times$ & $\times$ & \checkmark \\
    \rowcolor{lightGreen}
    \cellcolor{morelightGreen} User-Satellite Link Channel Model & $\times$ & $\times$ & \checkmark & \checkmark & \checkmark & \checkmark & \checkmark & $\times$ & \checkmark & \checkmark \\
    \rowcolor{lightGreen}
    \cellcolor{morelightGreen} OTFS-Based SatCom & $\times$ & $\times$ & $\times$ & $\times$ & $\times$ & $\times$ & $\times$ & $\times$ & \checkmark & \checkmark \\
    \rowcolor{lightGreen}
    \cellcolor{morelightGreen} Satellite Antenna Design & \checkmark & $\times$ & $\times$ & \checkmark & \checkmark & $\times$ & $\times$ & $\times$ & $\times$ & \checkmark \\
    \rowcolor{lightGreen}
    \cellcolor{morelightGreen} Multi-Satellite Systems & $\times$ & $\times$ & $\times$ & $\times$ & $\times$ & $\times$ & $\times$ & $\times$ & \checkmark & \checkmark \\
    \hline
  \end{tabular}
\end{table*}

In the preceding discussion, we highlighted several issues associated with terrestrial networks that can be addressed through the application of satellite and multi-satellite systems. We also acknowledged the issue of latency in satellite systems, which can be mitigated to a degree within the 5G framework.
Within this framework, we are eager to discuss a novel approach in SatCom: the concept of direct user-satellite communication. While specific constellations, such as Globalstar and Iridium, currently facilitate direct communications between satellites and users, such as satellite phones and messenger devices, these platforms do not extend to providing broadband internet services \cite{brandon2000key}.

In contrast, OneWeb and Starlink offer broadband internet connectivity using Very Small-Aperture Terminals (VSATs) \cite{sedin2020throughput}.
Unmodified handheld devices, such as cell phones, could potentially provide uninterrupted global connectivity if they could achieve a high-throughput connection with satellites. However, the capacity of this type of connection is restrained by the size and gain of the user's antenna, as well as power limitations at both the user and satellite ends.
One approach to mitigate these issues is to use larger antennas on the satellites themselves, a method exemplified by the experimental AST Bluewalker-3, which utilizes a $64\,\text{m}^2$ antenna \cite{n2yo}. Nevertheless, this approach is still in its testing phase and faces several challenges including concerns about space debris, high costs associated with implementation and launching, and potential disruption of astronomical observations.

An alternative solution could be to augment data transmission rates using multi-satellite Multiple-Input and Multiple-Output (MIMO) systems. This could potentially reduce the need for larger antennas and also enhance the data throughput within the same constellation \cite{omid2023spacemimo}.
One approach involves serving multiple satellites, without any collaboration among them, to only take the advantages of MIMO techniques.
Collaboration, on the other hand, involves the exchange of data or Channel State Information (CSI) between different satellites to facilitate tasks such as beamforming, detection, or precoding. By considering collaboration between satellites, the advantages of the aforementioned techniques can complement the MIMO technique, enhancing overall system performance.
For effective collaboration, Inter-Satellite Links (ISLs) play a crucial role, enabling seamless data and CSI sharing between satellites. ISLs come in various forms, such as Radio Frequency (RF), Terahertz (THz), or Free Space Optic (FSO) ISLs, depending on the specific requirements of the communication network. 
\textcolor{black}{It can also adopt a hybrid FSO/RF ISL configuration.}

\textcolor{black}{One of the significant concerns in Low Earth Orbit (LEO) SatCom is the Doppler effects due to the high speed of satellites. To address this issue, Orthogonal Time Frequency Space (OTFS)-based satellite systems offer a potential solution. OTFS modulation is a promising technology for SatCom, especially in LEO scenarios, as it effectively mitigates the severe Doppler effects in such high-mobility environments  \cite{10314509}. This issue will be widely discussed in our survey.
}

\begin{figure*}[t]
    \centering
\includegraphics[trim=3cm 4cm 1cm 4cm, clip, scale=0.44]{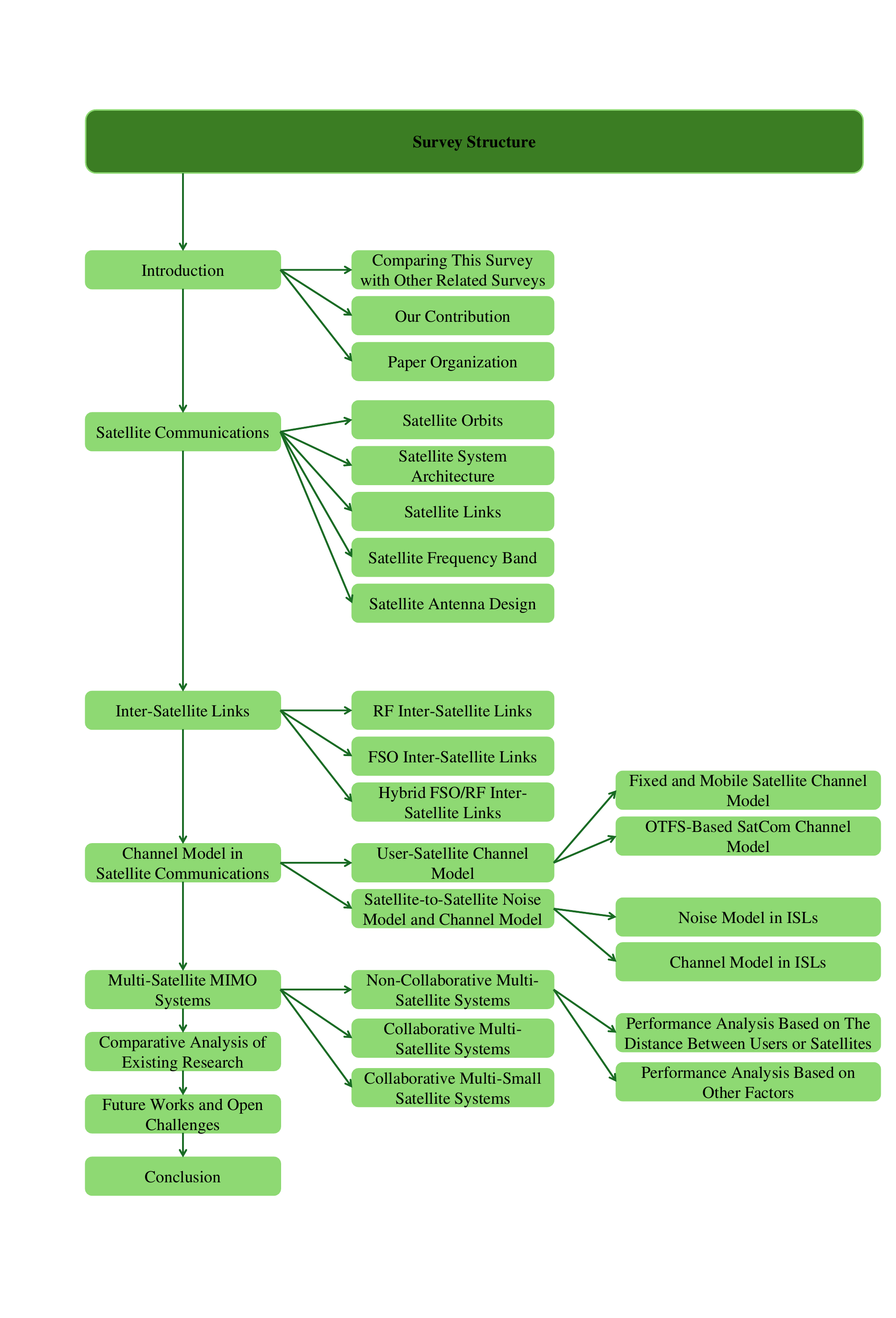}
\caption{Structure of the paper and topic classification}
\label{fig1}
\end{figure*}

\subsection{Comparing This Survey with Other Related Surveys}

\textcolor{black}{
Table \ref{tab1} presents a comparative analysis between our survey and other relevant surveys, with a focus on the main topics of our survey paper including "ISL Comparison", "ISL Channel Model", "User-Satellite Link Channel Model", "OTFS-Based SatCom", "Satellite Antenna Design", and "Multi-Satellite Systems". We explain below to show that although certain topics are addressed in other surveys, they are approached from a different perspective in our survey.}

\textcolor{black}{
User-satellite link channel model is explored in various survey papers, each approaching it from different perspectives. For instance, the authors of \cite{arapoglou2010mimo} examine the channel effect and the channel model of fixed and mobile satellite systems. 
The authors of \cite{prol2022position} extend the examination of channel models in mobile satellite systems, which was initially discussed in [18]. They begin by summarizing the channel effects observed in LEO SatCom. Following this, they introduce and discuss the different channel models employed in LEO SatCom scenarios.
Additionally, the authors of \cite{Baeza2022Overview} investigate the Land Mobile Satellite (LMS) channel model, and the authors of \cite{heo2023mimo} explore the multi-beam satellite channel model.
In our survey, we avoid delving into the details of the user-satellite link channel model, as this area has already been extensively covered. Instead, we focus on providing an overview of OTFS-based SatCom and exploring its channel model. While this topic is briefly mentioned in \cite{heo2023mimo}, the authors do not delve into its channel model.
The topic of ISL has been explored in \cite{al2022survey}, \cite{centenaro2021survey}, \cite{kodheli2020satellite}. However, these studies primarily compare two types of ISL: RF and FSO, discussing their advantages, disadvantages, and practicality. In our survey, we offer a concise summary of these qualitative aspects and pivot towards investigating the quantitative aspects of ISL, such as capacity, BER, or outage probability. Additionally, we review other studies that consider hybrid FSO/RF ISL to address the challenges associated with each type.
Furthermore, to comprehensively explore the topic of channel modeling, our survey paper also investigates the ISL channel model.
}

\textcolor{black}{
Within the domain of antenna design in SatCom, the authors of \cite{al2022survey} delve into the exploration of multi-beam and phased array antennas for NGSO satellites. Meanwhile, the authors of \cite{prol2022position} discuss the commonly used antenna types for LEO SatCom, and the authors of \cite{kodheli2020satellite} examine the advantages and disadvantages of active and passive antennas for satellites. 
In contrast, our focus is on addressing the challenges associated with using multi-element antennas to enable MIMO in SatCom, along with proposing solutions to overcome these challenges.
}
\textcolor{black}{
To provide a more detailed comparison in terms of multi-satellite systems topic, it is important to note that there are three scenarios for MIMO SatCom: single satellite with multi-beam, integrated satellite and terrestrial networks, and multi-satellite. These three scenarios are well-identified in \cite{heo2023mimo}.
While the authors of \cite{al2022survey} explore massive MIMO in NGSO satellites, their study does not address multi-satellite systems. Similarly, the authors of \cite{arapoglou2010mimo} focus solely on single or dual-satellite MIMO systems without delving into multi-satellite MIMO configurations.
Although \cite{heo2023mimo} is a valuable survey in the domain of MIMO SatComs, it does not extensively explore the multi-satellite MIMO scenario.
In contrast, the emphasis of our survey is on this scenario, which involves the use of multiple satellites. In addition, we explore direct mobile-to-satellite communications.
Furthermore, our survey categorizes multi-satellite systems into collaborative and non-collaborative types. Collaborative systems involve satellites sharing data via ISLs, while non-collaborative systems only relay signals.}

We present relevant references in both categories, providing readers with a comprehensive understanding of the multi-satellite MIMO scenario.
While direct communication between mobile devices and satellites is still in its early stages, our survey aims to be a valuable resource for those entering research in this field. By providing relevant references and insightful perspectives, our goal is to provide readers with a solid foundation as they delve into direct mobile-satellite communications.

\subsection{Our Contribution}

In summary, the main contributions of this paper are as follows:
\begin{itemize}
    \item \textcolor{black}{The objective of this survey is to explore the means of enabling direct communication between mobile devices and satellites through the utilization of multi-satellite MIMO systems. Consequently, 
    }
    this survey provides a comprehensive understanding of multi-satellite MIMO systems. 
    The use of multiple satellites for routing the data can act as a virtual MIMO system, taking advantage of MIMO techniques. As a result, they can play a crucial role in increasing data transmission rate (capacity) especially in the context of direct communication from handheld devices to satellites.
    
    \item We define collaboration between satellites as the exchange of data or CSI between them. In our study, we consider two cases: one where satellites do not collaborate and solely leverage the advantages of MIMO techniques to increase the channel capacity, and the other where they share CSI to employ additional techniques such as beamforming, precoding, or detection, which in turn improve the performance of the system.

    \item \textcolor{black}{We present a summary of antenna design specifically tailored for SatCom systems. We discuss the important considerations for designing satellite antennas and highlight various design techniques aimed at fulfilling these requirements. In particular, we focus on the deployment of multi-element antennas in SatCom configurations and address the obstacles involved in achieving optimal performance.
    }

    \item \textcolor{black}{ISLs can be established using either RF or FSO technologies, each with its own set of limitations. To address some of these drawbacks, hybrid FSO/RF ISLs emerge as a promising solution. Consequently, in our survey, we delve deeper into the exploration of Hybrid FSO/RF ISLs and investigate their advantages compared to both RF and FSO ISLs. We delve into research papers that employ hybrid ISLs to tackle the limitations of both individual transmission methods.}  

    \item \textcolor{black}{
       In LEO SatCom, the Doppler effect caused by satellite movement poses a significant challenge. OTFS modulation effectively tackles this issue by mitigating severe Doppler effects in dynamic environments. Additionally, in multi-satellite MIMO systems, varying signal reception delays occur due to significant distances between satellites. OTFS modulation resolves this by converting time-variant channels into time-independent ones in the Delay-Doppler (DD) domain. So, in the channel model section, we provide an in-depth examination of the channel model tailored for OTFS-based satellite systems.}
        
    \item  \textcolor{black}{
    ISLs play a vital role in facilitating collaboration between satellites in MIMO SatCom systems. However, it is important to acknowledge that ISLs are not ideal in practice and can introduce various challenges to signal transmission. Factors such as pointing error loss and noise significantly impact the performance of these links and must be accurately modeled and accounted for in system design. Consequently, our study delves deep into ISL channel models for both RF and FSO communications. Furthermore, within this section, we extensively analyze the noise models applicable to both RF and FSO ISLs.}

\end{itemize}

\subsection{Paper Organization}

The remaining sections of this survey are structured as follows:
\textcolor{black}{
To provide essential background information for those beginning to explore multi-satellite systems, 
}
section \ref{Satellite communications} explores SatComs, covering various aspects such as satellite orbits, architecture, frequency bands, antenna design, and different links used in SatCom.
\textcolor{black}{
ISLs play a vital role in ensuring efficient collaboration among satellites within a multi-satellite system.
}
This subject is addressed in Section \ref{Inter Satellite Links}, incorporating discussions on RF, FSO, and hybrid FSO/RF inter-satellite communication.
Section \ref{Channel model in satellite communications} focuses on the channel model in SatComs. It delves into both the user-satellite and satellite-to-satellite channels, offering insights into their characteristics and implications. Additionally, this section provides a review of the OTFS-based SatCom channel model 
Section \ref{Multi satellite MIMO systems} outlines our contributions to multi-satellite MIMO systems. It discusses both non-collaborative and collaborative strategies and also includes a focus on the collaboration of small satellites.
In Section \ref{Comparative Analysis of Existing Research}, a comparative analysis of relevant literature is conducted, showcasing simulation results. 
In Section \ref{FutureWorks}, we explore the plans and remaining challenges in the field of multi-satellite MIMO system research.
Lastly, Section \ref{Conclusion} concludes this survey, summarizing the key findings and insights.
To allow readers to easily follow the above-mentioned information we have included a visual representation of the paper's structure in Fig. \ref{fig1}. Furthermore, to assist readers in understanding the various abbreviations used throughout the paper, we have compiled a comprehensive list of acronyms in Table \ref{tab2}.

\begin{table*}
\centering
\caption{LIST OF ACRONYMS}
\label{tab2}
\begin{tabular}{ll|ll}
\rowcolor{morelightGreen}
\hline \textbf{Acronym \qquad\quad} & \textbf{Definition \qquad\qquad\qquad\qquad\qquad\qquad\qquad\quad\qquad} & \textbf{Acronym\qquad\quad} & \textbf{Definition \qquad\qquad\qquad\qquad\qquad\qquad\qquad\quad\qquad} \\
\rowcolor{lightGreen}
\hline
3GPP & 3rd Generation Partnership Project & LEO & Low Earth Orbit \\
\rowcolor{morelightGreen}
\hline
5G & Fifth Generation & LMS & Land Mobile Satellite \\
\rowcolor{lightGreen}
\hline
6G & Sixth Generation & LNA & Low Noise Amplifier \\
\rowcolor{morelightGreen}
\hline
APD & Avalanche Photodiode & LNB & Low Noise Blocks \\
\rowcolor{lightGreen}
\hline
AWGN & Additive White Gaussian Noise & LoS & Line of Sight \\
\rowcolor{morelightGreen}
\hline
BBSP & Best Beam Selection Policy & LTE & Long-Term Evolution \\
\rowcolor{lightGreen}
\hline
BER & Bit Error Rate & MEO & Medium Earth Orbit \\
\rowcolor{morelightGreen}
\hline
BPP & Binomial Point Process & MER & Modulation Error Ratio \\
\rowcolor{lightGreen}
\hline
BS & Base Station & MIMO & Multiple-Input and Multiple-Output \\
\rowcolor{morelightGreen}
\hline
CCI & Co-Channel Interference & MIMOSA & MIMO Channel Model for Mobile Satellite Systems \\
\rowcolor{lightGreen}
\hline
CDMA & Code Division Multiple Access & mMIMO & massive MIMO \\
\rowcolor{morelightGreen}
\hline
CFO & Carrier Frequency Offset & MMSE & Minimum Mean Square Error \\
\rowcolor{lightGreen}
\hline
CMMC & Collocated Massive MIMO Connectivity & MSE & Mean Square Error \\
\rowcolor{morelightGreen}
\hline
CNR & Carrier-to-Noise Ratio & MU-MIMO & Multi-User MIMO \\
\rowcolor{lightGreen}
\hline
CSI & Channel State Information & NASA & National Aeronautics and Space Administration \\
\rowcolor{morelightGreen}
\hline
DD & Delay-Doppler & NLoS & Non-Line of Sight \\
\rowcolor{lightGreen}
\hline
DE & Deterministic Equivalence & non-i.i.d. & independent but not necessarily identically distributed \\
\rowcolor{morelightGreen}
\hline
DgNB & Donor gNodeB & NTN & Non-Terrestrial Networks \\
\rowcolor{lightGreen}
\hline
DGS & Defected Ground Structures & OTFS & Orthogonal Time Frequency Space \\
\rowcolor{morelightGreen}
\hline
DM-MIMO & Distributed Massive MIMO & PDF & Probability Density Function \\
\rowcolor{lightGreen}
\hline
DMMC & DM-MIMO Connectivity & QoS & Quality of Service \\
\rowcolor{morelightGreen}
\hline
DSDP & Dual Satellite Dual Polarization & RAN & Radio Access Network \\
\rowcolor{lightGreen}
\hline
DSS & Distributed Satellite System & RF & Radio Frequency \\
\rowcolor{morelightGreen}
\hline
DSSP & Dual Satellite Single Polarization & RN & Relay Node \\
\rowcolor{lightGreen}
\hline
EBG & Electromagnetic Band Gap & RTT & Round Trip Time \\
\rowcolor{morelightGreen}
\hline
EO & Earth Observation & SAGIN & Space Air Ground Integrated Network \\
\rowcolor{lightGreen}
\hline
EO-SAR & Earth Observation and Synthetic Aperture Radar & SatCom & Satellite Communication \\
\rowcolor{morelightGreen}
\hline
ES & Earth Station & SCS & Sub-Carrier Space \\
\rowcolor{lightGreen}
\hline
FDMA & Frequency Division Multiple Access & SISO & Single Input Single Output \\
\rowcolor{morelightGreen}
\hline
FoA & Formation of Array & SNR & Signal-to-Noise Ratio \\
\rowcolor{lightGreen}
\hline
FSO & Free Space Optic & SRR & Split Ring Resonators \\
\rowcolor{morelightGreen}
\hline
GCS & Ground Control System & SS & Spread Spectrum \\
\rowcolor{lightGreen}
\hline
GEO & Geostationary Earth Orbit & SSC & Single Satellite Connectivity \\
\rowcolor{morelightGreen}
\hline
GNSS & Global Navigation Satellite System & TDD & Time Division Duplex \\
\rowcolor{lightGreen}
\hline
GPS & Global Positioning System & TDMA & Time Division Multiple Access \\
\rowcolor{morelightGreen}
\hline
HEO & Highly Elliptical Orbit & THz & Terahertz \\
\rowcolor{lightGreen}
\hline
HPA & High Power Amplifier & TM-TC & TeleMetry and TeleCommand \\
\rowcolor{morelightGreen}
\hline
HTFS & High Throughput Fractionated Satellite & TT \& C & Tracking, Telemetry, and Command \\
\rowcolor{lightGreen}
\hline
i.i.d. & independent and identically distributed & UE & User Equipment \\
\rowcolor{morelightGreen}
\hline
IoT & Internet of Things & UHF & Ultra-High Frequency \\
\rowcolor{lightGreen}
\hline
ISL & Inter Satellite Link & VSAT & Very Small-Aperture Terminals \\
\rowcolor{morelightGreen}
\hline
ISS & International Space Station & XPD & Cross-Polarization Discrimination \\
\rowcolor{lightGreen}
\hline
ITU & International Telecommunication Union & & \\
\hline
\end{tabular}
\end{table*}

\section{Satellite communications}
\label{Satellite communications}
In this section by introducing key concepts, technologies, and challenges in SatCom, we aim to make the transition to multi-satellite MIMO SatCom smoother and more comprehensible for our readers. Our discussion will enhance the reader's understanding of satellite orbits, satellite system architectures, satellite links, frequency bands utilized in SatComs, and satellite antenna design.

\subsection{Satellite Orbits:}
The altitude of a satellite, which signifies the distance between the satellite and the Earth's surface, determines the speed at which the satellite orbits around the Earth. Orbital characteristics, such as eccentricity (deviation from a perfect circle) and inclination (tilt relative to the equator), further influence the satellite's orbit type \cite{giambene2018satellite, wang2022ure}. 
One type of satellite orbit is GEO, where satellites reside at an altitude of $35,786\,\text{km}$ above the equatorial plane and exhibit near-zero inclination. These satellites synchronize their orbital period with the Earth's rotation, resulting in a 24-hour cycle. From an observer's standpoint, GEO satellites seem stationary at a fixed longitude, although some slight drift may occur over time. GEO satellites offer coverage over about one-third of the Earth's surface, excluding the polar regions, and their networks provide ubiquitous coverage to 99\% of the world's populated areas \cite{Nsengimana,pratt2019satellite}.
Despite their expansive reach, GEO satellites encounter a higher propagation path loss, necessitating the use of larger antennas and higher transmit powers to compensate for signal degradation. Moreover, GEO satellites exhibit a higher latency compared to constellations at lower altitudes, which makes them less suitable for time-sensitive services. The latency from GEO satellites to Earth is approximately $256\,\text{ms}$, whereas certain Medium Earth Orbits (MEOs) whose height below $10000\,\text{km}$ and LEO satellites have latency around $71\,\text{ms}$. Consequently, recent interest has shifted towards MEO and LEO satellites as more viable alternatives \cite{kodheli2017integration}.

MEO satellites operate at altitudes ranging from $2000\,\text{km}$ above the Earth's surface to the same height as GEO satellites, with variable orbital periods typically falling between 2 to 24 hours \cite{lutz2012satellite}.
LEO satellites occupy altitudes spanning from $400\,\text{km}$ to $2000\,\text{km}$ above the Earth's surface. These satellites orbit the Earth at a speed of about $7.8\,\text{km/s}$, completing a full rotation every 1.5 to 2 hours \cite{maini2011satellite}. Due to their rapid movements, LEO satellites swiftly change their positions relative to an observer on the ground, remaining visible for only brief periods before moving out of sight \cite{lin20215g}. 
The document \cite{5gntnwhitepaper} provides additional information and study findings concerning the duration of satellite visibility to the User Equipment (UE).
Various factors impact this visibility, including the orbit altitude, beam footprint radius, satellite velocity, UE velocity, the relative movements between the UE and the satellite, and whether a moving or fixed-beam scenario is assumed.
These factors collectively determine the length of time the satellite remains within the UE's Line of Sight (LoS).
Currently, numerous services are offered by different LEO satellite systems. These include broadband connectivity (e.g., Iridium, OneWeb, and Starlink), IoT applications \cite{centenaro2021survey} (e.g., Hiber, Myriota), Earth Observation and Synthetic Aperture Radar (EO-SAR) applications (e.g., Iceye, HawkEye) \cite{prol2022position}. These LEO systems play pivotal roles in various sectors, such as telecommunications, IoT, and Earth Observation (EO).

\begin{figure}[t]
    \centering
\includegraphics[trim=1cm 1.5cm 2cm 2.5cm, clip, scale=0.22]{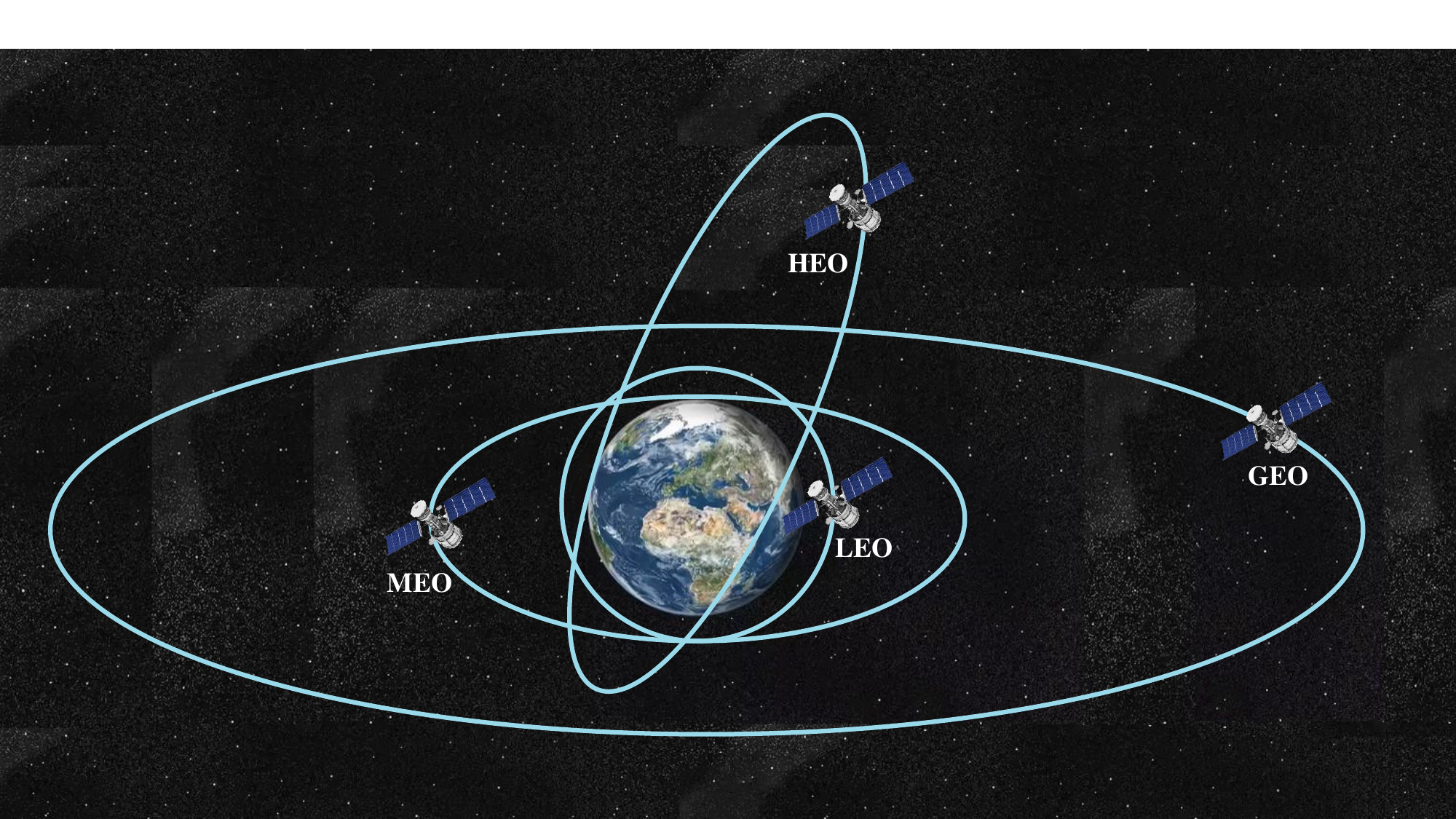}
\caption{Types of satellite orbits}
\label{fig2}
\end{figure}

In SatCom services, circular orbits are commonly employed. However, there are cases where highly eccentric elliptical orbits, known as Highly Elliptical Orbits (HEO), are utilized.
HEO orbits exhibit a significant characteristic: satellites move faster when closer to the Earth and slower when farther away. They enable satellites to move at a slower pace, remaining at high altitudes over ground sites in high-latitude regions. As a result, elliptical orbits are well-suited for SatComs, as they can maintain stable connections for extended periods. 
In these orbits, the eccentricity ranges between 0.63 and 0.71. This specific eccentricity range ensures that the apogee point (the farthest point from Earth) provides sufficient coverage over the polar zone, making HEO orbits ideal for monitoring polar regions effectively and continuously \cite{trishchenko2016multiple, roshanreview}. Nations like Russia rely heavily on HEO to provide coverage over polar and near-polar regions.
In contrast, GEO is not suitable for serving high latitudes due to their lower elevation above the horizon, which limits their coverage in these areas. Notable examples of inclined HEOs include Molniya and Tundra orbits. An illustration of different satellite orbits is shown in Fig. \ref{fig2}.

Non-Geostationary Earth Orbit (NGEO) satellites, including LEO, MEO, and HEO satellites, possess distinct features that differentiate them from conventional GEO satellites.
Notably, NGEO satellites typically have reduced propagation delay (except HEO in their apogee), smaller physical size, lower power consumption in both user and satellites and minimized signal losses \cite{al2022survey,richharia2014mobile}.
MEO and LEO are often deployed in satellite constellations. This approach is necessary because the coverage area offered by a single satellite in these orbits is limited and moves as the satellite travels at a high angular velocity to remain in its orbit. To ensure continuous coverage over a larger area including high-latitude regions, many MEO or LEO satellites are required \cite{babich2019nanosatellite,wood2003satellite}. This stands in contrast to GEO satellites, which operate at a much higher altitude, moving at the same angular velocity as the Earth's rotation, and thus provide permanent coverage over a vast area using just one satellite.
The Doppler effect is also an issue for the NGEO satellite systems. The movement of satellites introduces additional Doppler effects, especially for satellites at lower altitudes. This requires careful consideration in the design of UE, especially for wide-band links.

To mitigate significant Doppler shifts caused by satellite movements in NGEO SatCom systems, a technique called pre-compensation can be used for forward link signals. This involves applying a time-varying frequency offset that tracks the Doppler shift to the reference frequency of the forward link. The goal is to ensure that the forward link signals received at a specific reference point within the spot beam appear to have no Doppler shift. However, it's important to note that the Doppler shift differences vary across different locations within the spot beam and change over time \cite{lin20215g}.

\begin{figure}[t]
    \centering
\includegraphics[trim=4.3cm 3cm 0cm 1cm, clip, scale=0.33]{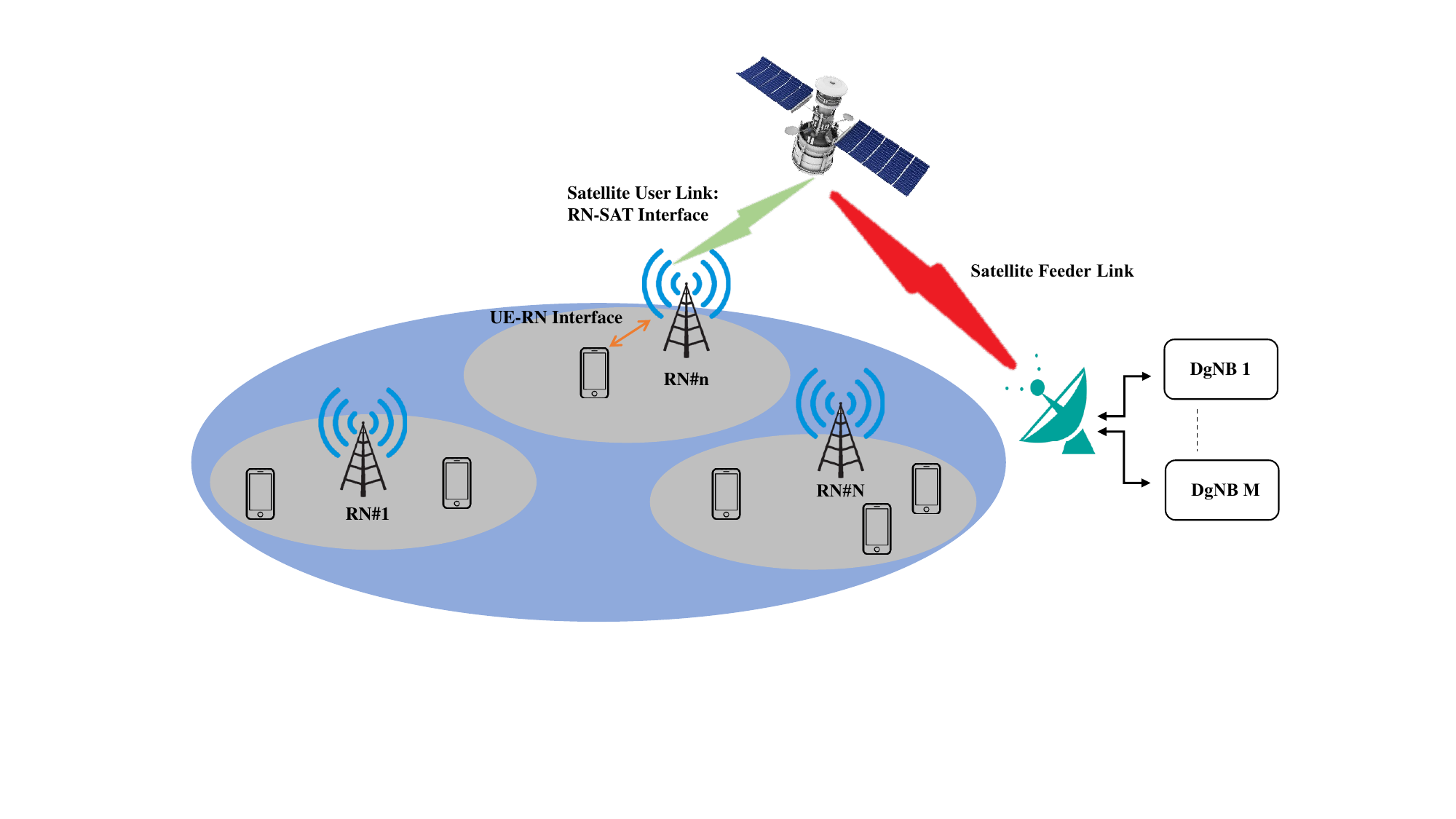}
\caption{g-Sat-RN Architecture \cite{kodheli2017integration}}
\label{fig3}
\end{figure}

The authors of \cite{kodheli2017integration} consider a scenario in which some UEs are connected to 5G Relay Nodes (RNs). A gateway station links to the satellite, providing access to the Donor gNodeB (DgNB), which connects the RNs to the 5G core network. 
To gain a clearer understanding, the illustration of their system model can be found in Fig. \ref{fig3}.
There are two types of Doppler shifts to consider: one experienced by the UE due to its motion relative to the RN (handled by RNs), and another caused by satellite movement (to be investigated).
To handle large Doppler shifts in the satellite channel, they suggest equipping RNs with Global Navigation Satellite System (GNSS) receivers to estimate the satellite's position. This compensation significantly reduces the Doppler shift. However, some residual Doppler remains due to estimation errors.
Their research reveals that the largest residual Doppler occurs at a 90-degree elevation angle with a fixed estimation error. To handle larger estimation errors, they propose increasing the Sub-Carrier Space (SCS). Therefore, a waveform with high SCS is preferable in the RN-DgNB interface, ensuring better performance in this scenario.
Regarding Doppler compensation, in addition to work in \cite{kodheli2017integration}, Krondorf et al. provide an illustrative algorithm for logic-efficient recursive Doppler rate estimation, offering a practical example in this regard. This algorithm can assist in handling Doppler shifts accurately in SatCom systems\cite{krondorf2020logic}.

\subsection{Satellite System Architecture:}
\label{Satellite Architecture}
The satellite system comprises three essential components: the space segment, the control segment, and the ground segment. The space segment consists of a constellation of active and spare satellites responsible for communication and data transmission. On the other hand, the control segment includes ground-based facilities, such as Tracking, Telemetry, and Command (TT$\textit{\&}$C) stations, that monitor and control the satellites, manage onboard resources, and ensure efficient utilization. The ground segment comprises Earth Stations (ESs) that establish communication links with the satellites and facilitate data exchange. These stations vary in size, ranging from small to large, depending on service requirements \cite{maral2020satellite}. This ground segment acts as a vital interface connecting the satellite system with end users, enabling connectivity and communication services \cite{elbert2014satellite}.

The space segment can be divided into two main components: the payload and the platform. 
The platform is the physical structure or satellite bus that houses essential subsystems and components for the satellite's operation in space.
The payload has two common types: the transparent payload and the regenerative payload. The transparent payload, also known as a "bent pipe" payload, focuses on amplifying carrier power and down-converting frequencies. The functionality of the transparent payload, as defined in 3GPP TR 23.737 \cite{3GPPTR23737}, involves the following process.
Initially, electromagnetic waves propagated from the Earth's surface are received by a satellite's antenna. These waves are then converted into an electric signal. The signal is subjected to channel filtering and amplification through a Low Noise Amplifier (LNA).
Subsequently, the signal undergoes frequency conversion, which prepares it for transmission.
Finally, the transformed signal is delivered to a transmitting antenna using a High-Power Amplifier (HPA). The transmitting antenna generates a reconditioned electromagnetic wave directed towards the Earth's surface, where the receiving gateway is located.
The transparent architecture's advantage lies in its independence from the radio waveform, eliminating the need for changes in the space-borne station. However, it faces drawbacks including amplified noise, vulnerability to jamming attacks, and the absence of ISL connections for traffic steering\cite{angeletti2006beam, mazzali2015board}.

In contrast, the regenerative payload involves demodulating uplink carriers, allowing for on-board processing and routing of information through on-board switching at base-band \cite{evans1999satellite}. Frequency conversion within a satellite is achieved by modulating onboard-generated carriers at the downlink frequency. These modulated carriers are then amplified and forwarded toward the desired downbeam destination, facilitating efficient communication \cite{evans1999satellite}. According to \cite{3GPPTR38821} the use of a regenerative payload is necessary for initial implementations of ISLs ensuring reliable and efficient communication between satellites over long distances in space.

\subsection{Satellite Links:}
\begin{figure}[t]
    \centering
\includegraphics[trim=1cm 0cm 0cm 0cm, clip, scale=0.22]{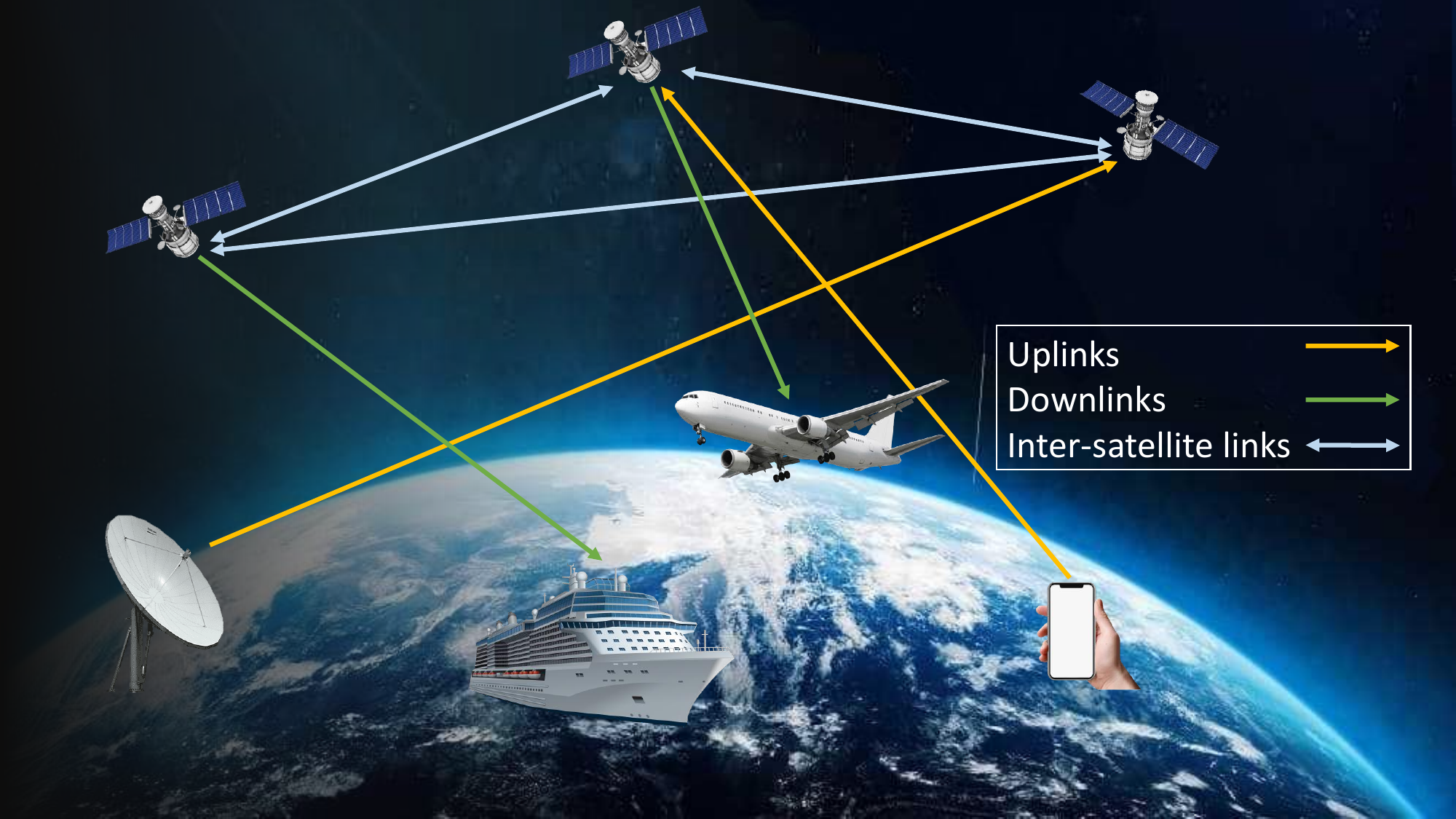}
\caption{Different Satellite Links}
\label{fig4}
\end{figure}

A communication link between a transmitter and a receiver involves using a modulated carrier, which can take the form of either an RF or an optical signal. There are three main types of links: uplink, downlink, and ISL which are shown in Fig. \ref{fig4}. In the uplink signals are transmitted from a user to a satellite, in the downlink, signals are sent from a satellite to a user, and the ISL establishes connections between the satellites. Uplinks and downlinks primarily utilize RF-modulated carriers, while ISLs can employ either RF or optical carriers. These carriers are modulated with baseband signals that carry the necessary information for communication \cite{maral2020satellite}. More details about ISLs will be described later in section \ref{Inter Satellite Links}.

\subsection{Satellite Frequency Band:}
SatComs utilize a broad range of frequency bands, covering frequencies from $1\,\text{GHz}$ to $50\,\text{GHz}$ \cite{kodheli2020satellite}.
Table \ref{tab3}  displays the main frequency bands commonly used in SatCom.
\begin{table}
\centering
\caption{DIFFERENT FREQUENCY BANDS USED IN SATCOM \cite{kodheli2020satellite}}
\label{tab3}
\begin{tabular}{p{3cm}|p{3cm}}
\hline
\rowcolor{morelightGreen}
\textbf{Frequency Band} & \textbf{Frequency} \\
\hline
\rowcolor{lightGreen}
L-Band & $1-2\,\text{GHz}$\\
\hline
\rowcolor{morelightGreen}
S-Band & $2-4\,\text{GHz}$\\
\hline
\rowcolor{lightGreen}
C-Band & $4-8\,\text{GHz}$\\
\hline
\rowcolor{morelightGreen}
X-Band & $8-12\,\text{GHz}$\\
\hline
\rowcolor{lightGreen}
Ku-Band & $12-18\,\text{GHz}$\\
\hline
\rowcolor{morelightGreen}
K-Band & $18-26\,\text{GHz}$\\
\hline
\rowcolor{lightGreen}
Ka-Band & $26-40\,\text{GHz}$\\
\hline
\rowcolor{morelightGreen}
Q/V-Band & $40-50\,\text{GHz}$\\
\hline
\end{tabular}
\end{table}
Within this range, different frequency bands are chosen based on factors such as climate conditions, types of services, and types of users. These bands are commonly denoted by single letters, distinguishing between lower frequencies (L, S, X, and C-bands) and higher frequencies (Ku, K, Ka, and Q/V-bands). The L-band is utilized by radio navigation systems like Global Positioning System (GPS) or Galileo, while the S-band finds applications in weather radar, surface ship radar, and communication between National Aeronautics and Space Administration (NASA) satellites and the International Space Station (ISS) and Space Shuttles.

Both the L and S-bands are also used for TT$\textit{\&}$C purposes. Furthermore, the frequency range between $2-2.3\,\text{GHz}$ is shared by space research, space operations, and EO satellite services.
It is worth noting that the lower frequency bands have limited bandwidth availability, making them a valuable and costly resource.
As a result, the C and Ku-bands are mainly used for SatComs, particularly for TV broadcasting.
However, the congestion of these bands has become a significant concern, necessitating a shift towards the Ka-band. The Ka-band offers a significantly larger signal bandwidth compared to the C-band and the Ku-bands, enabling higher data rates and improved capacity \cite{guan2019review, perez2019signal}. Nonetheless, it is important to consider that Ka-band systems are more vulnerable to adverse weather conditions compared to the Ku-band and especially the C-band systems. On the other hand, higher frequency bands facilitate the use of smaller antenna sizes, promoting the adoption of compact and efficient multi-antenna systems \cite{kodheli2020satellite}.

\textcolor{black}{\subsection{Satellite Antenna Design}}
\textcolor{black}{Different types of antennas utilized in LEO satellites are explored in \cite{prol2022position}, offering valuable insights into the application of each antenna type. The use of active antennas for enabling MIMO techniques in LEO satellites is investigated in \cite{al2022survey}, while \cite{kodheli2020satellite} examines active antenna architectures across GEO, MEO, and LEO satellite systems. Additionally, passive antennas for GEO satellites are analyzed based on their advantages and disadvantages. The paper also discusses strategies for reducing the number of active elements in an active antenna to lower costs and complexity, addressing challenges such as the grating lobe problem and proposing solutions.
However, design challenges and protocols for satellite antennas have not been thoroughly surveyed yet. In this subsection, we offer an overview of antenna design for SatCom, focusing on multi-element antennas to leverage the benefits of the MIMO technique.}

\textcolor{black}{The design of satellite antennas necessitates consideration of several key factors. Some of them include concerns about antenna gain, size, coverage of multiple bands, and, in the case of MIMO antennas, isolation between elements. Needless to say, achieving high gains is essential to compensate for the path loss during long-range signal transmissions, and the compactness of the antenna enables it to fit within the limited space available on satellites. Additionally, the capability to operate across multiple frequency bands is a significant advantage for satellite antennas and enables the integration of satellites with various communication technologies.
Against this background, several design techniques are employed to meet these requirements. These include the use of planar patch antennas, as detailed in \cite{8819861,8474859,ta2019planar,hati2017novel}, which offer a balance of compactness and efficiency. Reflector and lens antennas, discussed in \cite{119577,8852830}, are known for their high gain and directivity, but their large size is a disadvantage. Slot antennas, discussed in \cite{9737521,tubbal2019high}, are another approach valued for their adaptability and performance in different frequency bands. Additionally, transmit and reflect array antennas, referenced in \cite{arya2019large} are popular choices due to their ability to form directed beams and operate efficiently across various frequencies.}

\textcolor{black}{The MIMO antennas used in satellite communications often feature a planar structure with various potential shapes. A critical factor in enhancing MIMO performance for satellite applications is the maximization of isolation between radiating elements \cite{9737521}. The main challenging part of designing a compact MIMO antenna is achieving a high isolation level, preferably greater than $20\ \text{dB}$, between the two adjacent antenna elements without affecting the scattering, radiation, and diversity performances.
To address this, recent research has focused on several decoupling techniques. These include the use of Electromagnetic Band Gap (EBG) substrates, as discussed in \cite{wu2018quad}, which are effective in reducing mutual coupling. Defected Ground Structures (DGS), referenced in \cite{8010273}, offer another method for enhancing isolation. The use of metamaterials and metasurfaces, discussed in reference \cite{9234406}, has also gained attention due to their unique electromagnetic properties. Additionally, techniques employing Split Ring Resonators (SRR) and parasitic resonators, cited in references \cite{alharbi2022multi} and \cite{8226753} respectively, have been reported. These approaches have shown promise in achieving the desired high isolation levels in MIMO antennas for SatComs.}\\

\section{Inter-Satellite Links}
\label{Inter Satellite Links}
The primary focus of this survey is to examine and analyze the use of multi-satellite MIMO systems as a solution for addressing the link budget limitations in direct connectivity scenarios, where satellites within such systems are interconnected through ISLs.

ISLs serve as a means of communication between satellites, facilitating the transmission of high-speed data and, if desired, controlling the information related to the satellite navigation and the flight control functions such as TeleMetry and TeleCommand (TM-TC) \cite{radhakrishnan2016survey}. Alternatively, each satellite could establish a connection with a BS. However, this approach presents two significant challenges. Firstly, it is probable that not all satellites would have a direct LoS to the BS. Secondly, the individual connections from each satellite to the BS would encounter substantial path loss, thereby hindering the closure of the link budget. Moreover, compared to simple uplinks and downlinks between a satellite and its station on the ground, ISLs enable satellite-to-satellite relays, thus reducing the need for stations on the ground. This technology also permits the use of frequencies that are quickly attenuated in the atmosphere, making the link undetectable and unjammable from the ground. It offers real-time or near-real-time communication capabilities for command, control, communication, and information processing \cite{muri2012survey}. Consequently, it is more advantageous for the satellites to engage in collaboration among themselves, rather than relying solely on connections to the BS.

\begin{figure}[t]
    \centering
\includegraphics[trim=7cm 2cm 2cm 4cm, clip, scale=0.45]{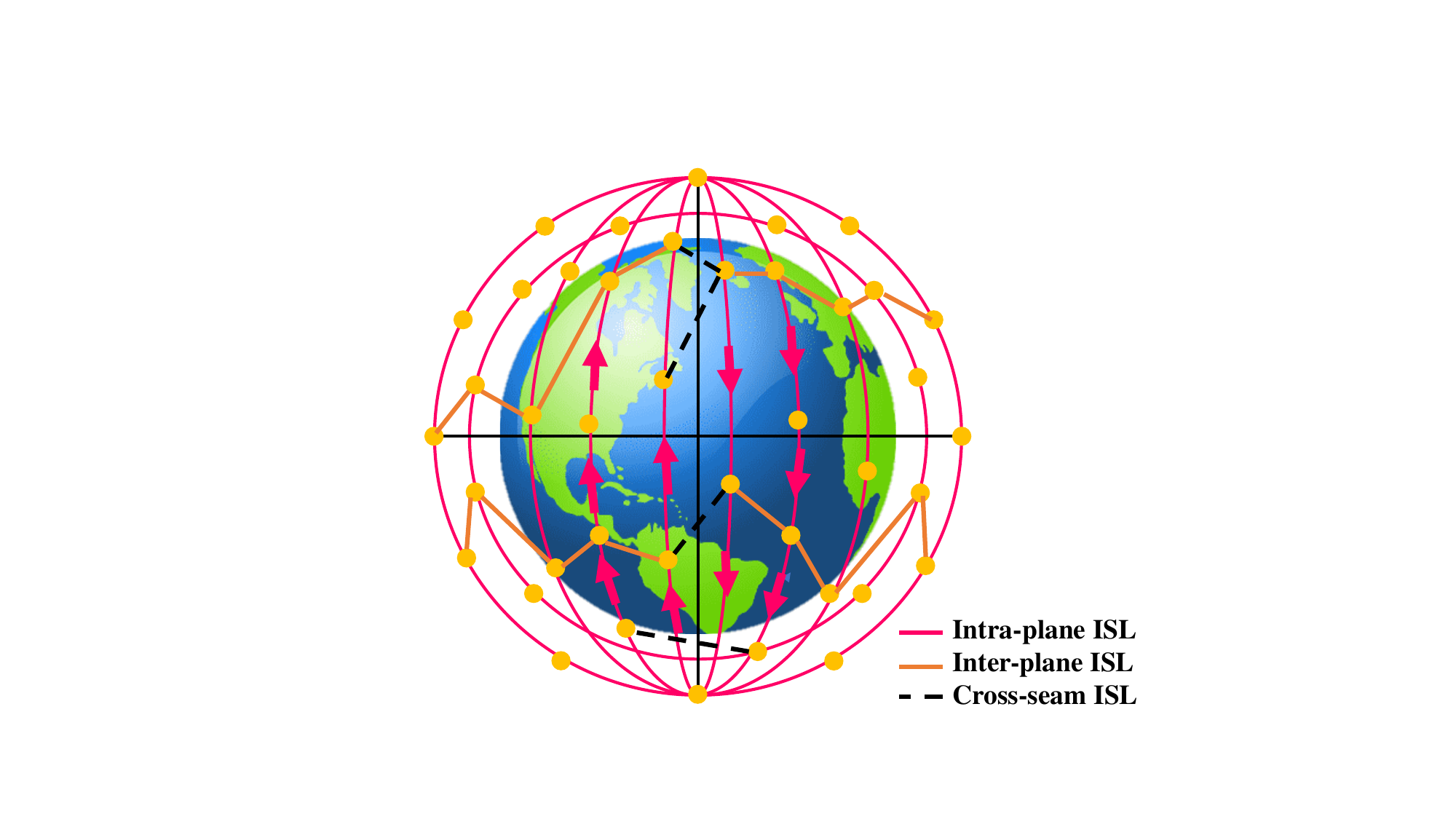}
\caption{Different definition of ISLs\cite{leyva2020leo}}
\label{fig5}
\end{figure}

ISLs can be categorized into two types: intra-plane and inter-plane. 
Intra-plane ISLs allow satellites within the same orbital plane to communicate with each other. These links are stable because neighboring satellites in the same orbital plane maintain consistent distances, which is called intra-plane distance.
In contrast, inter-plane ISLs enable communication between satellites in different orbital planes. These links are more dynamic because satellites have varying velocity vectors, leading to changes in their relative positions.
In certain cases, satellites may orbit the Earth in nearly opposite directions, leading to significantly higher relative velocities between them. These ISLs connecting different orbital planes, referred to as cross-seam ISLs, experience noticeable Doppler effects. Satellites and their ISL connections can be arranged in different configurations. For more detailed information about these ISL constellations and the related RF aspects, refer to \cite{leyva2020leo}. Fig. \ref{fig5} provides visual examples of different types of ISLs. 

Various works in the literature have discussed ISLs. Muri et al. provide a survey on the frequency, protocol, application, and orbit of satellite systems that have launched with ISLs until 2015\cite{muri2012survey}. 
In \cite{sun2010enabling}, the research is divided into three main parts that address different aspects of inter-satellite communication. 
The first part focuses on basic communication issues that are relevant to inter-satellite communication, including the choice of RF or optical links, frequency allocations, data rate, and power control mechanisms. 
The second part examines the inter-satellite communication network architectures and technologies and provides an overview of Frequency Division Multiple Access (FDMA), Time Division Multiple Access (TDMA), and Code Division Multiple Access (CDMA) for sharing links in Distributed Satellite Systems (DSS). 
The third part of the paper summarizes existing inter-satellite ranging methods that refer to the measurement or estimation of the distance between two or more satellites within a satellite system. By exploring these different aspects of inter-satellite communications, the authors aim to provide insights and recommendations for the development of a robust and efficient inter-satellite communications and ranging system for small satellites.
In \cite{chen2021modeling}, the authors consider and formulate some parameters that are crucial for the performance analysis of inter-satellite communications. These parameters include the distance between two satellites and the elevation and azimuth angles from one satellite to another.

ISLs in multi-satellite systems can be established using either wireless or wired connections. Typically, wired connections are utilized in satellite swarms and are called tethered formations.
Utilizing tethers, which establish a physical connection between swarm elements, offers substantial benefits in simplifying swarm operations.
The presence of a tethered structure significantly mitigates the impact of perturbations on the formation, ensuring greater stability. 
However, it is important to consider that the use of tethers introduces certain complexities. For instance, re-configuring the system in space becomes more challenging due to the presence of physical connections. It also presents mechanical challenges in designing a deployable structure that can efficiently reduce the swarm's size for launch purposes \cite{tuzi2023satellite}.
However, in scenarios where wireless connections are employed, ISLs have the flexibility to utilize various communication media such as RF, FSO, or hybrid FSO/RF links as their primary means of communication \cite{shah2017short, kaushal2015free}. 

\textcolor{black}{The authors of \cite{al2022survey} and \cite{kodheli2020satellite} thoroughly examine the advantages and disadvantages of RF and FSO ISLs. Additionally, the authors of \cite{al2022survey} delve into interference management in ISLs in detail. 
Furthermore, the authors of \cite{radhakrishnan2016survey} investigate the implementation of ISL, specifically focusing on small satellite communications.
In the following discussion, we survey RF and FSO ISLs from the perspective of system performance aspects. This expanded analysis covers topics such as throughput maximization and link budget calculations, resulting in a more comprehensive examination of ISLs.
Moreover, in our survey, we extend the analysis to include hybrid FSO/RF ISLs.}

\subsection{RF Inter-Satellite Links}
The first ISLs relied on microwave RF, operating in Ka-band and V-band frequencies. In the WRC19 report \cite{international2020world}, the International Telecommunication Union (ITU) has outlined specific frequency ranges for RF ISLs. These include ranges such as $14.8-15.35\,\text{GHz}$, $17.8-18.4\,\text{GHz}$, $22.55-23.55\,\text{GHz}$, $24.45-24.75\,\text{GHz}$, $25.25-27.5\,\text{GHz}$ and $66-71\,\text{GHz}$.
Some satellite constellations such as Iridium have deployed RF ISLs for their operations.
Iridium satellites use ISLs to provide satellite calls and digital data services, with each satellite supporting up to four simultaneous ISLs at carrier frequencies ranging from $22.55-23.55\,\text{GHz}$, and a data rate of $25\,\text{Mbps}$ \cite{pratt1999operational, braun2021satellite}. 

An instance of utilizing ISLs is exemplified in \cite{jia2017collaborative}. In situations where a satellite with a large amount of data to transfer in the downlink has a short contact time with the ES, the scheduling algorithm may not be able to downlink all the data to the ES.
To deal with this issue, the authors of \cite{jia2017collaborative} propose offloading the data among satellites using ISLs before the data is sent to the ES to maximize the overall system throughput.
By adjusting the amount of data each satellite carries to downlink to the ES, the satellite can fully utilize its contact time with the ES.
The authors of \cite{jia2017collaborative} compare their method with a method that does not use ISL for inter-satellite data transfer.
Simulation results show that downlink data transmission throughput can be significantly increased by using ISL data offloading, with the throughput reaching close to 100\% in many cases.
However, it is assumed that the communication on ISLs and between satellites and the ES is error-free.

Some researchers are studying the potential of the THz band in SatComs. This band which lies between RF and optical frequencies, offers a natural solution for ISLs to achieve ultra-high throughput with its broad spectrum.
The authors of \cite{nie2021channel} characterize the propagation channel in the THz spectrum for ISL at various frequencies, including $1\,\text{THz}$, $2\,\text{THz}$, $10\,\text{THz}$, with a bandwidth of $2\%f_c$. Furthermore, to make comparisons, they examine other frequency bands, including $26\,\text{GHz}$, $120\,\text{GHz}$, $300\,\text{GHz}$, and the optical band at $193\,\text{THz}$ ($1550\,\text{nm}$).
They categorize the channels into near-Earth and deep-space channels according to the satellite’s altitude and analyze their propagation features separately, and the effects of orbital perturbations on link throughput and outage probability. 
Simulation results show that the THz band provides significantly higher throughput compared to the Ka-band at $0.026\,\text{THz}$ and has a much lower outage probability compared to the optical link at $193\,\text{THz}$ (wavelength of $1550 \, \text{nm}$) under random orbital perturbation effects. 
THz band antenna arrays offer a wider beam with a higher tolerance for beam misalignment compared to optical-based solutions, and THz band ultra-wide bandwidth supports multi-Gbps of throughput, promising higher spectral efficiency and resolution.

The authors of \cite{ding2016analysis} focus on the capacity of a THz inter-satellite communication system, specifically analyzing the performance of high-gain reflective antennas. 
They derive the design parameters for high-gain reflective antennas that would enable the system to achieve a capacity of more than $100\,\text{Gbps}$ over short distances ($1\,\text{km}$) and $2.5\,\text{Gbps}$ over long distances ($100\,\text{km}$).

\subsection{FSO Inter-Satellite Links}

When evaluating the performance of an FSO ISL, it is essential to take into account both the internal and the external parameters of the system. 
The internal parameters are specific to the system and include factors such as transmission bandwidth, optical power, wavelength, types of lasers used, receiver's field of view, receive lens diameter, and receiver's sensitivity.
Conversely, the external parameters are dependent on the environmental conditions and include factors such as atmospheric attenuation, scintillation effects, pointing error losses, and deployment distance.
The authors of \cite{kaur2019analysis} primarily focus on the internal parameters and investigate the combined effect of different operating wavelengths, detector types, and pointing errors on the performance of an FSO ISL. 
They optimize the link performance by analyzing the BER and the quality factor of the received signal under different scenarios. 
Additionally, they investigate the performance of the ISL under different modulation formats and data rates for LEO and MEO distances.

In the research presented in \cite{chaudhry2021laser}, a classification for FSO ISL within a constellation is provided based on satellite location and link duration.
Based on satellite location within a constellation, two main types of FSO ISLs exist:
\begin{itemize}
    \item Intra-orbital plane FSO ISL: this type is established between two satellites positioned within the same Orbital plane.
    \item Inter-orbital plane FSO ISL: this type is created between satellites situated in different Orbital planes.
\end{itemize}
Additionally, FSO ISLs can be divided into two types based on their duration:
\begin{itemize}
\item Permanent FSO ISLs: these links are established between satellites for an extended period.
\item Temporary FSO ISLs: these links serve a specific duration or temporary purpose.
\end{itemize}

In \cite{maamar2016study}, the authors examine the link budget and communication performance of FSO ISLs. The received power at the receiver satellite is obtained by $P_R=P_TG_TG_RL\tau_T\tau_R,$
in which $P_T$, $G_T$, $G_R$, $L$, $\tau_T$ and $\tau_R$ stand for the transmitted power, the transmitter antenna gain, the receiver antenna gain, the free space path loss, the transmitter efficiency and the receiver efficiency respectively.
The transmitter and the receiver antenna gain obtained by $G_T=\frac{32}{\theta_T^2}$, and $G_R=(\frac{\pi D_R}{\lambda})^2$, respectively. Where $D_R$ is the receiver antenna diameter, $\lambda$ is the optical wavelength and $\theta_T$ is the transmitter divergence angle given by $\theta_T=1.22(\frac{\lambda}{D_T})$ \cite{djordjevic2010channel}, in which $D_T$ is the transmitter antenna diameter.
Moreover, the free space path loss is denoted by $L=(\frac{\lambda}{4\pi d})^2$, in which $d$ is the communication distance. All $D_R$, $D_T$, $d$ and $\lambda$ are measured in meters.   
Simulation results suggest that increasing the transmit power or receiver antenna diameter can compensate for increased communication distances.
Additionally, as the communication distance increases, the data rate decreases, thus, for long distances, a higher power is needed to enhance the data rate.

\subsection{Hybrid FSO/RF Inter-Satellite Links}

\textcolor{black}{Optical wireless SatCom links address the issue of limited data rates seen in RF and microwave links. FSO inter-satellite connections provide high data rates thanks to their extensive bandwidth availability, which benefits a variety of communication types, such as massive machine communication.
Despite its benefits, FSO communication faces challenges like beam instability, pointing errors, free-space loss, and weather conditions. While inter-satellite communication eliminates concerns about free-space loss and weather, the problem of pointing errors remains a challenge. Promising solutions such as hybrid FSO/RF ISLs can address the issue of FSO ISL 
\cite{9865150}.}

In \cite{liu2020inter}, using the combined advantages of FSO and RF ISLs, a hybrid FSO/RF ISL network is presented to improve system performance.
Specifically, FSO ISLs are used for providing high-speed communication, while RF ISLs provide flexible inter-satellite measurements.
To address the challenge of scheduling the two types of ISLs for optimizing the system performance, a compatible scheduling scheme is proposed. Simulation results demonstrate that the proposed hybrid ISL network outperforms an RF-only ISL network, with higher throughput (over $125\,\text{Mbps}$) and lower inter-satellite communication delay (less than $10\,\text{s}$).
\textcolor{black}{
In \cite{apoorva2020best}, authors suggest using hybrid ISLs to enhance communication reliability. However, this approach introduces new challenges, such as switching between different physical layers at the satellite's transmitter. In their work, they propose an innovative hybrid FSO/RF satellite system. To improve ISL reliability, they develop a new policy called the Best Beam Selection Policy (BBSP) and implement FSO and RF switching. They further investigate BBSP's performance by deriving expressions for outage probability, average spectral efficiency, and average BER. When it comes to physical layer switching, they compare error probabilities between RF and FSO links and determine the Signal-to-Noise Ratio (SNR) threshold for efficient switching to RF.}

\section{Channel model in satellite communications}
\label{Channel model in satellite communications}

Multi-satellite MIMO refers to a scenario where there are multiple satellites instead of just one satellite. Each satellite can have either a single antenna or multiple antennas. On the other hand, satellite MIMO typically involves a single satellite equipped with multiple antennas.
In the case of a multi-satellite MIMO system, there are multiple links between each satellite and the users in which each link's channel characteristics should be modeled.

\textcolor{black}{
In this section, we investigate channel models in both user-satellite link and satellite-to-satellite link. Moreover, we discuss about noise model in ISLs. Furthermore, in the section on user-satellite channel model, we briefly mention different types of channel models that can be used in SatCom. However, we do not provide detailed explanations of them, as these channel models have been investigated in previous surveys. Instead, we aim to extract higher-level insights and key takeaways regarding satellite MIMO systems by concentrating on examining the channel models for OTFS-based SatCom and ISL, which, to the best of our knowledge, have not been surveyed before.}

\textcolor{black}{ISLs play a crucial role in facilitating synchronization and data sharing among satellites for collaboration. Understanding the channel model in ISLs is crucial for designing, optimizing, and operating Space MIMO systems effectively. It ensures reliable and high-capacity communication in satellite networks because channel models help predict communication system performance. By accurately modeling channel characteristics, we can estimate factors like signal degradation, interference, pointing error, and others that affect communication quality.
Moreover, another issue in SatCom is the Doppler effect, which poses a notable challenge in LEO SatCom due to the rapid movement of satellites in their orbits. In response to this challenge, OTFS modulation has recently emerged as a promising solution for LEO SatCom. Therefore, comprehending OTFS-based SatCom systems is also imperative in the context of Space MIMO.}

\subsection{\textcolor{black}{User-Satellite Channel Model}}

\textcolor{black}{\textcolor{black}{In SatCom, channels can be broadly classified as fixed and mobile satellite channels, representing channel models within the context of Fixed Satellite Systems and Mobile Satellite Systems respectively \cite{hofmann2016spatial}.
Fixed satellite channels are used for communication with stationary users, while mobile satellite channels are employed for communication with users in motion or mobile platforms.} In this subsection, a brief explanation of fixed and mobile satellite channel models is provided. However, for more detailed information about each channel model, we direct the reader to Table \ref{tab4} and the references mentioned therein. Table \ref{tab4} offers a comprehensive list of related references for each channel model for satellite-to-UT link. Following that, the channel model for OTFS-based SatCom systems is clarified.}

\subsubsection{Fixed and Mobile SatCom Channel Model}

\textcolor{black}{Survey \cite{heo2023mimo} focuses on investigating channel attenuation and the channel model for multi-beam satellites with reflector or UPA antennas. 
Moreover, in survey \cite{kodheli2020satellite}, the classification of channel models into fixed and mobile satellite systems is thoroughly explained. 
In \cite{Baeza2022Overview}, the authors provide a comprehensive overview of 13 different NGEO channel models. These models take into account diverse aspects such as orbits, frequency bands, UE, use-case scenarios, and features specific to each scenario. The authors have gone a step further by conducting a detailed classification, simplifying the process of comparing and distinguishing between these models.}

\textcolor{black}{As a summary of these 3 survey papers, user-satellite channels are classified as fixed and mobile channels. Fixed channels, model communication between satellites and stationary ground users, often using the pure LoS model, as $h_{m,n}=\frac{c}{4\pi f r_{m,n}}e^{-j\frac{2\pi f}{c}r_{m,n}},$
where $r_{m,n}$ is the distance between the $n$-th transmit antenna and the $m$-th receiver antenna, $c$ stands for the speed of light ($c=3\times10^8\,m/s$), and $f$ denotes the carrier frequency.
Conversely, in mobile channel situations where users are in motion, there are various models. One of them is the generic fast-fading model specified in 3GPP, as documented in \cite{melleinwhitepaper}. This model comprises 12 steps that provide a detailed explanation of how the channel is generated. Another widely referenced channel model in SatCom is the flat fading model which is a non-frequency selective channel model. They include multi-state with LOO distribution, Rician, and Nakagami-$m$ fading model. The selection of the appropriate model depends on the specific conditions and assumptions of the study.
In what follows, we further survey the literature on channel measurement and channel generation techniques.}

\begin{table*}
\centering
\caption{SATELLITE CHANNEL MODEL CLASSIFICATION}
\label{tab4}
\begin{tabular}{c|c|c}
\hline
 & \cellcolor{lightGreen} \textbf{Reference} & \cellcolor{lightGreen} \textbf{Channel model} \\
\hline
 \textbf{Channel Model in Fixed Satellite Systems} & \cellcolor{morelightGreen} {\cite{storek2020multi, schwarz2009channel, li2021capacity, li2021analysis, schwarz2019mimo, dou2014cooperative, liolis2007multi, zhang2021distributed}} & \cellcolor{morelightGreen} {Frequency flat fading channel model with pure LoS propagation} \\
\hline
 \multirow{6}{*}{\textbf{Channel Model in Mobile Satellite Systems}} & \cellcolor{lightGreen} {\cite{3GPPTR38901v14, adeyemo2012simulation}} & \cellcolor{lightGreen} {Generic fast fading model\footnotemark} \\
\hhline{~--}
 & \cellcolor{morelightGreen} {\cite{3GPPrel15, prieto2010versatile,omid2023spacemimo, juan2021time}} & \cellcolor{morelightGreen} {Two-state model with a Loo distribution within each state} \\
\hhline{~--}
  & \cellcolor{lightGreen} {\cite{fontan2001statistical, fontan1997complex, fontan2007consolidation, burzigotti2008dvb, bai2019channel}} & \cellcolor{lightGreen} {Three-state model with a Loo distribution within each state} \\
\hhline{~--}
 & \cellcolor{morelightGreen} {\cite{peng2021channel, karasawa2021multistate}} & \cellcolor{morelightGreen} {Four-state model with a Loo distribution within each state} \\
\hhline{~--}
 & \cellcolor{lightGreen} {\cite{li2022performance, tjhung1999fade, okati2022nonhomogeneous}} & \cellcolor{lightGreen} {Nakagami-$m$ fading model} \\
\hhline{~--}
  & \cellcolor{morelightGreen} {\cite{bai2017cooperative, hofmann2017multisatellite, ramamurthy2016mimo, li2022uplink, abdelsadek2021future, abdelsadek2022distributed, schwarz2017multiple, omid2023oncapacity, loo1985statistical, corazza1994statistical, abdi2003new}} & \cellcolor{morelightGreen} {Rician fading channel model} \\
\hline
\end{tabular}
\end{table*}

LMS-MIMO networks can be obtained using multiple satellites with single polarization antennas or by a single satellite with dual polarization antennas. The authors of \cite{king2011empirical} present a straightforward approach for developing an empirical stochastic dual circular polarized LMS-MIMO narrow-band channel model. The model is designed to accurately reflect channel characteristics in both LoS and NLoS scenarios. To achieve this, they employ an integration of large-scale fading generation, Markov chain generation, and small-scale fast fading generation. 
The model is validated using data collected from a suburban/rural area in an S-band three-lined road measurement campaign, with a carrier frequency of $2.45\,\text{GHz}$ and a $200\,\text{MHz}$ bandwidth. 
Specifically, the authors use the model to simulate a $2\times2$ MIMO system with dual circular polarized antennas at both ends. The results demonstrate good accuracy in both LoS and NLoS cases, as evaluated by comparing measured data to simulated values for the Rice factor, Cross-Polarization Discrimination (XPD), and correlation. 
Overall, the model's simplicity and ability to adjust the key parameters make it well-suited for conformance testing of satellite MIMO applications.

The main focus of \cite{burkhardt2014mimosa} is on how to generate an LMS channel model within the MIMO Channel Model for Mobile Satellite Systems (MIMOSA) project. The project has generated a large database that represents different environments and system configurations for modeling LMS channels. 
The authors have developed a modular modeling approach within MIMOSA, which allows for a detailed characterization of various parameters of the propagation channel and antenna arrays of a MIMO system.
The LMS channel is modeled by considering different fading effects, such as very slow fading, small-scale fading, and large-scale fading. 
The observed signal is a superposition of these three effects.
Additionally, the authors validate their proposed method by comparing the results of re-simulating a track of channel sounder measurements to the actual measurement data. 
They find that the simulation results match well with the delay profile and LoS power, although there is a slight deviation in the received power for NLoS. 
Overall, the authors' approach provides a detailed and flexible method for generating LMS channel models within the MIMOSA project, which can be used for various system configurations and environments.

\subsubsection{\textcolor{black}{OTFS-Based SatCom Channel Model}}

\textcolor{black}{One of the significant challenges in LEO SatCom is the Doppler effect, which results from the high mobility of these satellites in their orbits. To address this challenge, OTFS modulation has recently emerged as a promising choice for LEO SatCom.}
\textcolor{black}{
Here, we start by introducing OTFS, followed by a brief explanation of the channel model used in OTFS modulation. Then, we present a basic channel estimation process, drawing on the findings from the works cited in \cite{8671740,8424569,7925924,8377159,10268002}.} 

\textcolor{black}{OTFS modulation is a promising technique for SatCom, particularly in LEO scenarios, due to its ability to effectively mitigate the severe Doppler effects inherent in such high-mobility environments. OTFS modulation excels in LEO SatComs because it operates in the DD domain, allowing it to average out the rapid channel dynamics caused by satellite movement. By doing so, it overcomes the challenges posed by Doppler frequency shifts without the need for complex approaches \cite{10043628,10261131,10314509}}.
\textcolor{black}{OTFS modulation presents several significant advantages. First, OTFS can seamlessly integrate with existing Orthogonal Frequency Division Multiplexing (OFDM) systems, ensuring architectural compatibility with technologies like Long-Term Evolution (LTE). Its fundamental concept revolves around the transformation of the time-varying multipath channel into a two-dimensional channel in the DD domain. By employing equalization in this domain, OTFS ensures that all symbols transmitted within a frame encounter uniform channel gains, which enhances reliability.
Notably, the channel's reduced variability in the DD domain enhances the practicality and robustness of OTFS compared to other methods, thereby diminishing the overhead and complexity associated with physical layer adaptation.
Moreover, the compact representation of the DD channel in OTFS allows for efficient packing of reference signals, a crucial requirement for supporting the extensive antenna arrays used in massive MIMO applications \cite{9849120}.
In essence, OTFS emerges as a practical modulation scheme that addresses critical challenges in wireless communication, offering improved compatibility, reliability, and efficiency \cite{7925924}.
Note that in SatComs, the DD domain becomes more sparse due to the limited number of paths between ground terminals and satellites, thus reducing the complexity of OTFS detection even further.}

\begin{figure*}
    \centering
    \includegraphics[trim=3cm 7cm 3cm 6cm, clip, scale=0.65]{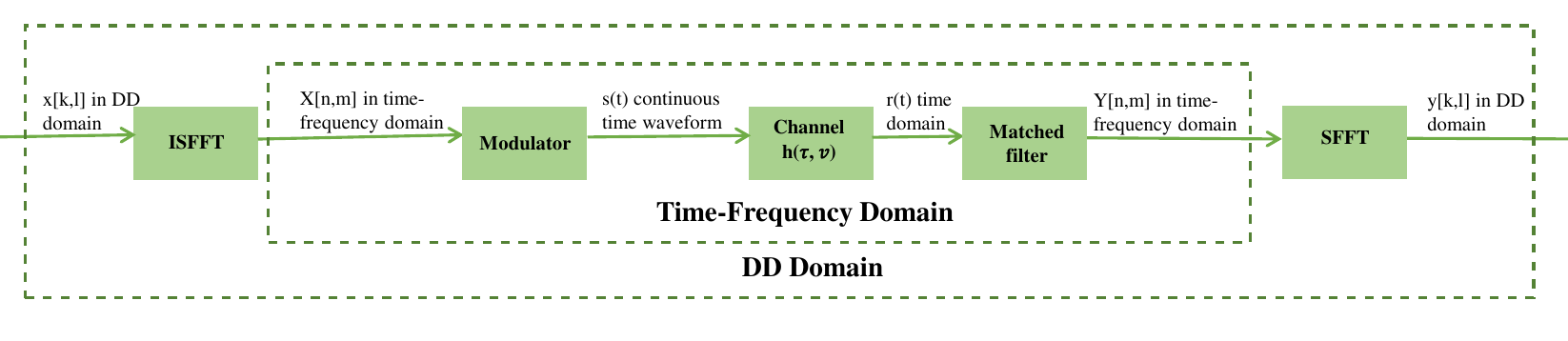}
    \caption{\textcolor{black}{OTFS system.}}
    \label{fig6}
\end{figure*}

\textcolor{black}{In OTFS, at first, the time-frequency domain is quantized into a grid by sampling the time and frequency at intervals $T$ seconds and $\Delta f$ Hz, as $\Omega = \{(nT,m\Delta f),n=0,..., N-1,m=0,..., M-1\}$. The values of $M$ and $N$ can be chosen based on the system characteristics and requirements \cite{8424569,10268002}. The DD plain is converted into an information grid, as $\Gamma=\{(\frac{k}{NT},\frac{l}{M\Delta f}),k=0,..., N-1,l=0,..., M-1 \}$ with $\frac{1}{NT}$ and $\frac{1}{M\Delta f}$ describing the quantization steps of the DD domain.
Now, a set of $NM$ information symbols $x[k,l]$, $k=0,..., N$, $l=0,..., M$, which are chosen from a QAM modulation alphabet, are arranged on the DD grid $\Gamma$. These symbols are converted into time-frequency domain symbols $X[n,m]$ using the inverse symplectic finite Fourier transform (ISFFT), and then, by using a transmit waveform, a time-frequency modulator converts those samples into a continuous time waveform $s(t)$.
The signal $s(t)$ is transmitted over a time-varying channel with complex baseband channel impulse $h(\tau,v)$ where $\tau$ and $v$ represent the delay and Doppler, respectively. Because the number of channel reflections is usually limited, a sparse representation of the channel can be given as $h(\tau,v)=\sum_{i=1}^Ph_i\delta(\tau-\tau_i)\delta(v-v_i)$,}
\textcolor{black}{where $P$ is the number of propagation paths between the transmitter and the receiver, $\delta(.)$ is a Dirac delta function, and $\tau_i$, $v_i$ and $h_i$ stand for the delay, Doppler, and gain of the $i$th path. The delay and Doppler taps for the $i$th path are given by $\tau_i=\frac{l_{\tau_i}}{M\Delta f}$ and $v_i = \frac{k_{v_i}}{NT}$, respectively. Here $l_{\tau_i}$ and $k_{v_i}$ represent the index of the delay and Doppler taps. At the receiver side, first, a matched filter computes the cross-ambiguity function using a receiver waveform, and the resulting signal is sampled at $t=nT$ and $f=m\Delta f$ instances where $n=0,..., N-1,m=0,..., M-1$. This process converts the received signal in continuous time $r(t)$ into a discrete time-frequency sample $Y[n,m]$. Finally, the receiver applies the symplectic finite Fourier transform to obtain the received symbols in the DD domain as $y[k,l]$. This entire procedure is shown in Fig. \ref{fig6}.}

\begin{figure*}
    \centering
    \includegraphics[scale=0.55]{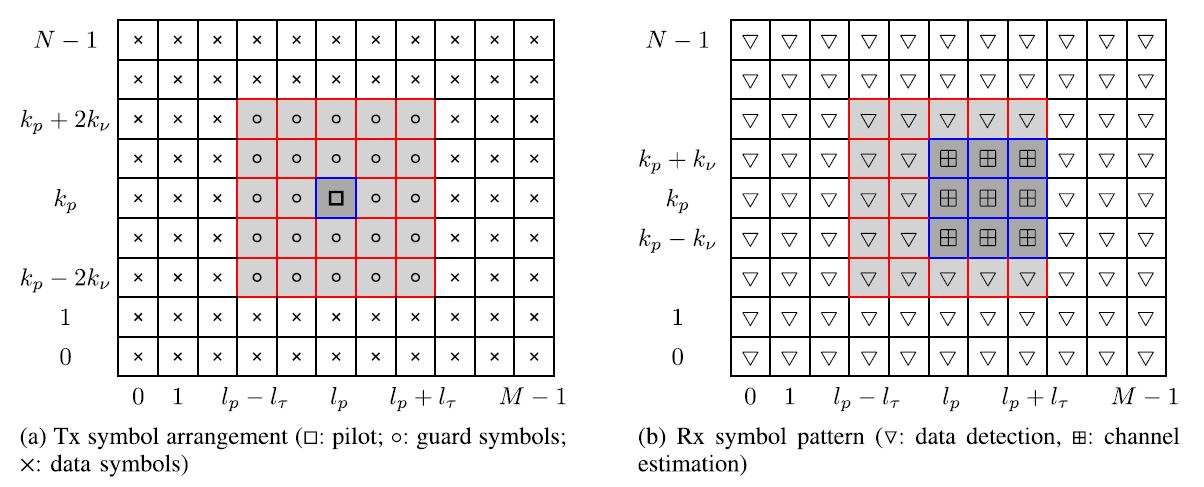}
   \caption{\textcolor{black}{DD domain pilot arrangement in OTFS modulation \cite{8671740}.}}
    \label{fig7}
\end{figure*}

\textcolor{black}{Now, focusing on the channel response, we see that in $h(\tau,v)$ only the gains of the propagation paths need to be estimated, which are invariant during the OTFS frame. Also, note that among the $M\times N$ DD grid of data, only a few paths exist that have their specific delays and Dopplers. This means that the channel is sparse, especially in SatCom, where LOS is dominant. Considering that the maximum delay among paths is $\tau_{max}$ and the maximum doppler is $v_{max}$, the index of the delay tap with $\tau_{max}$ delay is $l_{\tau}$ and the index of the Doppler tap with $v_{max}$ Doppler shift is $k_{v}$. The values of $k_{v}$ and $l_{\tau}$ are far less than the data grid sizes of $M$ and $N$. As for pilot transmission, we do not need to send a pilot in all $M\times N$ slots in the DD domain; we only need to send a pilot in one of the DD frames. After the detection, the value of the pilot can be detected in all of the paths, so the gain of the paths can be easily calculated. However, to avoid interference between the pilot and the data, some guard frames need to be placed around the pilot frame. To be more specific, we need to allocate $l_{\tau}$ guard frames on the delay axis and $2k_{v}$ guard frames in the Doppler axis around the pilot frame to avoid interference. Here is an example based on \cite{8671740} for a better understanding of this procedure. Assuming that the pilot is sent in the frame with the delay index tap $l_p$ and the Doppler index tap $k_v$, Fig. \ref{fig7}(a) shows the symbol and pilot arrangement for the transmitter. In that case, the received symbol pattern is shown in Fig. \ref{fig7}(b), where we see that the data detection and channel estimation are performed with no interference. Note that channel estimation in OTFS is another exciting research area that has been the focus of many papers, such as \cite{10138432,9785832} and references therein.}

\footnotetext{Fast fading and frequency selective model are used interchangeably in 3GPP Release 15 \cite{3GPPrel15}.}

\subsection{\textcolor{black}{Satellite-to-Satellite Noise Model and Channel Model}}

\textcolor{black}{In this section, we examine the noise and channel model associated with both RF/THz and FSO ISLs for a clearer understanding.}

\subsubsection{\textcolor{black}{Noise Model in ISLs}}
\textcolor{black}{In FSO and RF/THz communication systems, thermal noise is typically treated as Additive White Gaussian Noise (AWGN) \cite{apoorva2020best}.
Thermal noise arises from random electron fluctuations in a resistor due to temperature variations.
It is described as a Gaussian random variable with zero mean, where the variance is influenced by the load resistor $R_L$, the temperature $T_t$, and the noise figure of the electrical amplifier $F_n$, represented as $\sigma_{T_t}^{2}=4K_B\frac{T_t}{R_L}F_n B_n$ \cite{aladeloba2013optically}.}

\textcolor{black}{ISL channel is typically classified into two categories based on satellite altitude: near-Earth (with altitudes above $400\,\text{km}$ and limited by GEO at approximately $36000\,\text{km}$) and deep-space (satellites located beyond GEO).
In current SatCom network link budget analyses, noise temperature is often simplified as thermal noise, a valid assumption for near-Earth scenarios. However, temperature variations have a more significant impact on deep-space communication systems. As detailed in \cite{nie2021channel}, the solar brightness temperature contributes to the total noise temperature in deep-space. This contribution combines with cosmic microwave background noise and thermal noise temperature.}
\textcolor{black}{In RF and THz ISLs, thermal noise dominates the total noise. However, for FSO ISL employing an Avalanche Photodiode (APD) receiver, the total noise encompasses shot noise, thermal noise, and occasionally dark current noise (often neglected in practice).
So, the total noise is the sum of total thermal noise and APD shot noise.
Shot noise arises from random photon arrivals at the APD receiver and can be modeled as a zero-mean Gaussian random process \cite{magidi2021free}.
The overall APD noise can be accurately modeled as a zero-mean, signal-independent AWGN. The total variance is $\sigma_n^2=\sigma_{Sh}^{2}+\sigma_{ T_t }^{2}$ \cite{magidi2021free, 9108615}.}

\subsubsection{\textcolor{black}{Channel Model in ISLs}}

\textcolor{black}{In RF/THz ISLs, neglecting pointing error loss, attenuation, and free space path loss simplifies the channel model to a unity gain configuration, with ISL noise being the primary transmission factor. However, some researchers do consider pointing error loss for THz ISLs, making their channel model akin to that of FSO ISLs, where beam divergence and pointing accuracy significantly influence channel characteristics.}

\textcolor{black}{In the context of FSO channels, the optical irradiance is influenced by several factors, expressed as a product of three components: free space path loss, atmospheric attenuation, and pointing error loss as detailed in \cite{farid2007outage}.
The tropospheric effects relevant to user-satellite links, including attenuation and absorption due to rain, cloud, and fog, typically occur within the altitude range of 8–14 km. However, in near-Earth ISLs these effects can be disregarded, allowing us to consider the atmospheric attenuation as equal to 1 \cite{nie2021channel}.
Free space path loss is dependent on the distance between two satellites, $d$,  and the atmospheric attenuation factor, $\alpha$. It is calculated as $I_l=e^{-\alpha d}$.
However, certain authors argue that at the high altitudes of satellites, which are situated above the atmospheric layers, turbulence caused by atmospheric effects may not significantly impact the wireless channel. Consequently, they consider the free space path loss to be negligible at 0 dB/km \cite{9513570}.}
\textcolor{black}{The radial error of pointing jitter, denoted as $\theta$, follows a Rician distribution \cite{ma2012influence}.
For simplicity, it is common to consider the bias error angle as zero, which simplifies $\theta$ to follow a Rayleigh distribution \cite{9473712}.
The normalized pointing error loss can be calculated as $I_P(\theta)=\text{exp}\left(-\frac{2\theta^{2}}{\omega^{2}}\right)$ \cite{farid2007outage},
where $\omega$ represents the half-width divergence angle of the Gaussian beam.
So, the PDF of $I_P(\theta)$ can be derived from \cite{farid2007outage} as $f(I_p(\theta))=f(\theta) \left|\frac{d_{\theta}}{d_{I_p}}\right|=\frac{\omega^2}{4\sigma^2}I_p^{\frac{\omega^2}{4\sigma^2}-1}$.}

\section{Multi-satellite MIMO systems}
\label{Multi satellite MIMO systems}
\begin{figure}[t]
    \centering
\includegraphics[trim=4cm 2cm 2cm 2cm, clip, scale=0.38]{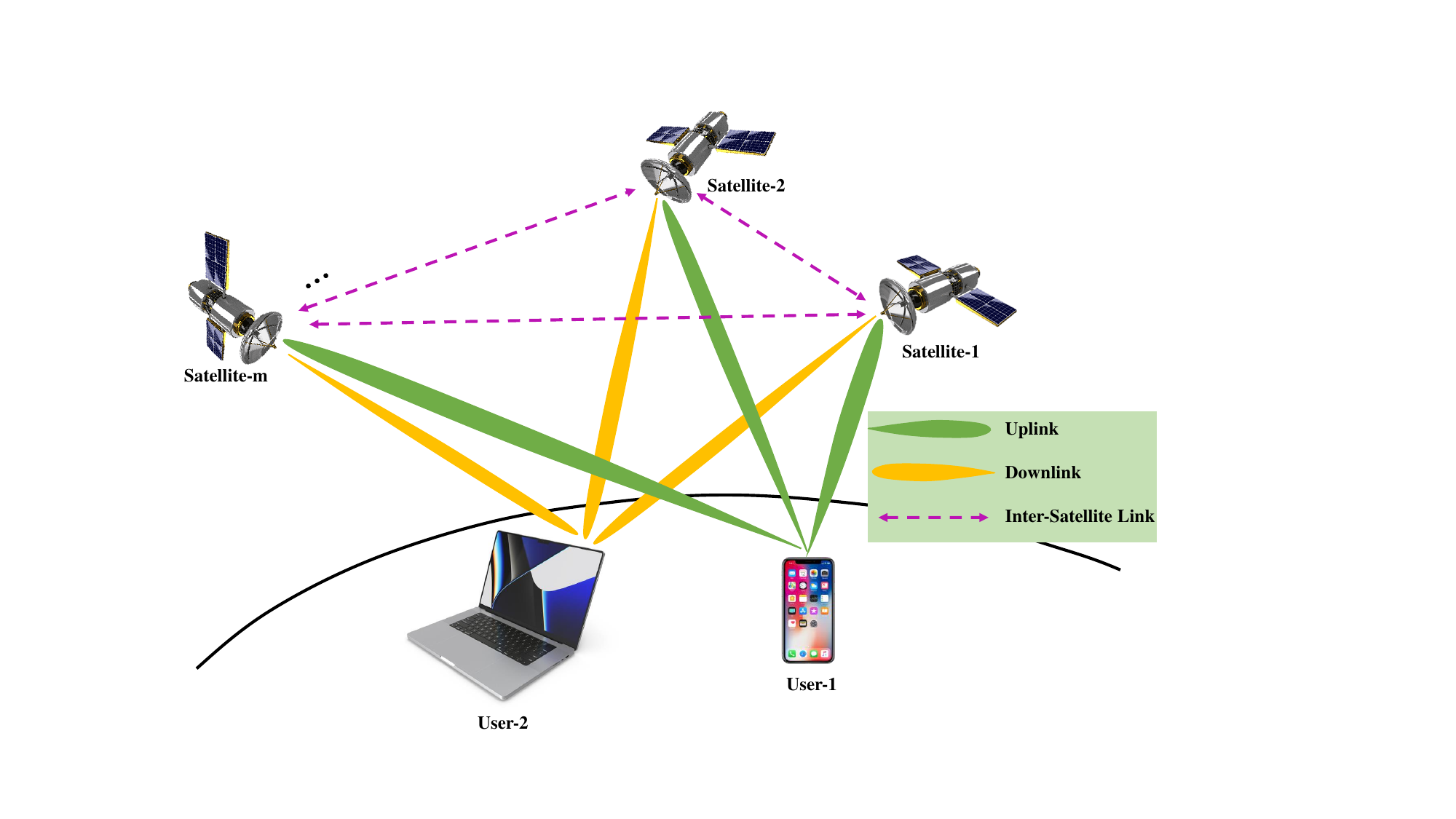}
\caption{Multi-satellite MIMO system}
\label{fig8}
\end{figure}

This section presents a summary of various works conducted in the field of multi-satellite MIMO systems. A visual representation in Fig. \ref{fig8} depicts a general system model of a multi-satellite MIMO system that incorporates collaboration among the satellites.

Satellite systems have captured the industry's attention for their potential to connect billions of mobile subscribers directly. Companies such as Apple \cite{apple2022iphone}, Huawei \cite{huawei2022satellite}, and Lynk \cite{lynk2022launches} are actively involved in utilizing satellite systems to provide basic emergency services. However, their ability to exchange data is limited, impacting the QoS compared to terrestrial networks. To enhance performance, it is necessary to address various challenges. In the context of traditional single-satellite communication systems, closing the link budget in direct connectivity scenarios requires equipping the payload with a larger antenna reflector and increasing power. A promising solution to overcome these challenges and improve performance is the adoption of a distributed satellite architecture for direct connectivity.

Multi-satellite architecture offers a valuable extension to single-satellite MIMO system, enabling cooperative operation among multiple satellites to serve multiple ground users and establish a virtual MIMO system. This technique enables the system to take advantage of MIMO transmission even when only a single antenna is available on satellites. This is particularly useful for situations where hardware or environmental restrictions prohibit the use of multiple antennas by satellites and ground users, thereby limiting the feasibility of employing single-satellite MIMO. \textcolor{black}{The case of a single satellite with multiple beams MIMO SatCom has been thoroughly investigated in survey \cite{al2022survey}.}
However, the application of MIMO techniques in SatCom, when using multiple antennas on one satellite, introduces a challenge related to the high spatial correlation between channels. As a result, the channel characteristics deviate from the typical MIMO scenario.
Furthermore, the benefit of multi-satellite diversity over polarization diversity is highlighted in \cite{horvath2006sat02}, as the former can lead to a capacity increase of $M$-fold, where $M$ is the number of satellites employed. In contrast, polarization diversity can only result in a transmission rate increase of up to a factor of two, as reported in \cite{horvath2006sat02}.

Unlike terrestrial mobile channels, satellite channels lack abundant scattering components, leading to increased correlation and reduced capacity in LoS dominant environments \cite{bai2018multi}. To overcome this issue, it becomes crucial to maintain a significant antenna spacing, either in space or on the ground \cite{arapoglou2010mimo}. Distributed multiple satellites can be employed to increase the distance between antennas in space, facilitating the compact arrangement of terrestrial antennas\cite{li2021capacity}.
In the study conducted by Sundaramoorthy et al. \cite{sundaramoorthy2010systematic}, various implementations of DSSs are discussed and compared. The paper highlights several configurations that can be realized in space, including formation-flying spacecraft, satellite constellations, fractionated spacecraft, and satellite swarms.
In the context of satellite constellations and formation flying spacecraft, the work in \cite{paul2011communication} provides an insightful definition and comparison followed by a description of fractionated satellites.

To classify relevant multi-satellite studies, we need to initially establish our understanding of what constitutes collaboration between satellites. While some researchers regard synchronization between satellites as a form of collaboration, we define collaboration between satellites as the exchange of data and/or CSI to perform tasks such as beamforming, precoding, or detection through data processing.

\subsection{Non-Collaborative Multi-Satellite Systems }
Most of the research on using the MIMO technology in SatComs primarily aims to improve the capacity using other methods, rather than investigating the potential benefits of satellite collaboration.
\textcolor{black}{In this section, we review MIMO SatCom studies that do not take into account satellite collaboration. In SatCom, addressing the challenge of correlation that results in capacity reduction necessitates maintaining a substantial antenna spacing, whether in space or on the ground \cite{li2021capacity}. Achieving the maximum capacity also requires determining the necessary spacing between user antennas.
Therefore, to emphasize studies centered on the distance between users or satellites in SatCom and to improve the organization of this subsection, we are dividing this section into two parts.   
In the first part, we explore research that primarily aims to maximize capacity by taking into account the distance between users or satellites. Towards the end of this section, we also discuss studies that center around random constellations where the distances between satellites are randomized.
In the second part, we examine studies that consider factors other than distance, influencing the performance of multi-satellite MIMO systems. These factors encompass the influence of rainfall on the link budget, various signal transmission methods, interference, Carrier Frequency Offset (CFO), Doppler shift, and more.}

\subsubsection{Performance Analysis Based on The Distance Between Users or Satellites}

\textcolor{black}{In this section, our attention is directed towards multi-satellite MIMO SatCom studies that operate without collaboration between satellites. These studies delve into system performance by examining the distances between users or satellite antennas.}

In \cite{ramamurthy2016mimo}, \textcolor{black}{the study involves deploying two transparent GEO satellites in the same orbit but with substantial orbital separation. These satellites are connected to two ground users located nearby. The study assumes that there is no collaboration between the satellites, and any necessary time delay compensation is managed at the receiver end. Furthermore, this study examines both the forward link and return link scenarios.}
Since Ultra-High Frequency (UHF) SatCom has limited bandwidth and capacity, this paper investigates the possibility of using MIMO to enhance the spectral efficiency without increasing the transmit power.
\textcolor{black}{The authors highlight the fact that} in the LoS SatCom channel, achieving spatial orthogonality requires a substantial antenna separation of several kilometers, either in space or on the ground. To overcome this requirement, the proposed solution involves deploying multiple satellites to achieve a large antenna separation (30 degrees) in space. This setup allows users on the ground to have closer antenna separations. 
\textcolor{black}{In their simulation findings, they illustrate how the spacing between user antennas impacts the achievable capacity. From the perspective of UHF users, the differences in uplink and downlink frequencies (in return and forward link respectively), result in distinct optimal antenna spacing requirements for the uplink and downlink directions. However, in their specific scenario where users are spaced $1\,\text{m}$ apart, they achieve a capacity that closely approximates the optimal capacity in both the return link and forward link scenarios.}

The work of Schwarz et al. \cite{schwarz2019mimo} examines a system similar to the one previously discussed, which takes into account both the effects of ground \textcolor{black}{antenna} and satellite antenna positions, as well as the perturbations of GEO satellites, in both uplink and downlink scenarios.
\textcolor{black}{
The authors conduct an analysis of channel capacity in two distinct scenarios. In one scenario, they account for satellite positioning errors, leading to a time-varying capacity gain, while in the other scenario, they omit this error consideration. Their findings reveal that even with relatively low GEO positioning accuracy, 85$\%$ of the optimal capacity can be attained in 90$\%$ of all observations, which represents a significant improvement compared to conventional Single Input Single Output (SISO) SatCom systems.}

\textcolor{black}{
Much like the study described in \cite{ramamurthy2016mimo}, the research in \cite{schwarz2017multiple}, explores the application of MIMO technology in UHF SatCom. However, in this study, the authors perform measurements in a scenario involving mobile satellite-to-car communication. They compare the measurement results with those obtained in a stationary scenario and their theoretical results in \cite{dantona2010uniform}. 
The measurement results match theoretical LoS predictions well. 
Their findings indicate that the maximum MIMO channel capacity can be attained in both stationary and mobile UHF scenarios. Furthermore, they confirm that the channel capacity is closely linked to the geometric arrangement of the MIMO antennas.}
\textcolor{black}{they conclude that in stationary scenarios, it is possible to optimize the UHF MIMO SatCom channel to reach its maximum MIMO channel capacity. Furthermore, in mobile scenarios, achieving nearly maximum MIMO capacity is feasible by avoiding the so-called "keyhole channel".}
\textcolor{black}{As described in} \cite{schwarz2008optimum}, the worst-case type of the MIMO channel which provides the minimum MIMO capacity, is referred to as the keyhole channel.

A user selection strategy has been proposed in \cite{dou2014cooperative} aiming to exploit the maximum MIMO capacity of a system with two GEO satellites that are closely separated in longitude. 
\textcolor{black}{In this study, the focus is on the uplink transmission from $N$ individual single-antenna users situated within a spot beam towards two GEO satellites. These satellites employ array antennas to create multiple beams, with each beam dedicated to a specific spot beam. As a result, for users within a spot beam, the satellites effectively function as a single antenna system.}
\textcolor{black}{Taking into account user cooperation in signal transmission,}
a maximum capacity criterion for cooperative transmission is derived based on the distance between two users. In this case $d_{opt}$ is the minimum equivalent distance of two users that can achieve optimum capacity. The equivalent distance of two selected cooperative users should be a multiple of $d_{opt}$. \textcolor{black}{Moreover, the authors of this study } consider a deviation degree of the distance between users from $d_{opt}$, denoted by $\Delta d$.
\textcolor{black}{In their user selection scenario,} a resource request for transmission is sent by the $i$th user to Ground Control System (GCS), and GCS finds another user in the spot beam that meets the maximum capacity criteria ($d=d_{opt}\pm\Delta d$). Selected users will then transmit signals simultaneously to satellites in the same frequency band and the same time slot. If GCS cannot find a suitable user, satellites and users will communicate in a non-cooperative manner, to avoid the extra cost of cooperation.
It is demonstrated that cooperative transmission rates oscillate periodically as the distance between two users increases. Furthermore, the achieved rate of cooperative and non-cooperative systems with satellite collocation is much higher than with one satellite with a high power level.

In \cite{bai2017cooperative} a system model similar to the work in  \cite{dou2014cooperative} is considered. 
\textcolor{black}{The main difference is that unlike \cite{dou2014cooperative} which solely looks at a LoS channel model, \cite{bai2017cooperative} considers severe weather conditions like rain and wet snow, which introduce multipath effects and result in a Rician fading channel. }
The authors demonstrate that improving the transmission rate is not only dependent on the user's position, similar to \cite{dou2014cooperative}, but also on the Rician K-factor.
In their analysis, the researchers assume an independent and identically distributed (i.i.d.) Gaussian distribution for K-factors. They proceed to plot the upper bound on the system's achievable transmission rate, considering various variances for the K-factors.
Based on their simulation results, the total system transmission rate reaches its maximum value in two cases, i) when the two users have pure NLoS channels ($\kappa_1=\kappa_2=0$), and ii) when the two users have pure LoS channels with good orthogonality.

\textcolor{black}{The previously mentioned works rely on deterministic constellations, meaning that the satellite positions are deterministic. However, these studies are not capable of offering a comprehensive analysis that can be applied to any arbitrary constellation because they are tied to specific network design parameters. Furthermore, traditional deterministic and location-based models, which have been used in the past for satellite system analysis, are usually restricted to simulation purposes. Consequently, some researchers employ stochastic geometry to assess the operational efficiency of LEO SatComs. This implies that the satellite positions within the constellation are no longer deterministic but instead distributed randomly.
As an example, a Binomial Point Process (BPP) is a way to describe how points are scattered within a limited area. Each point's location in this model is determined by a type of probability distribution called binomial distribution \cite{chiu2013stochastic}. When we think about the positions of LEO satellites flying at a certain height, we can use the BPP idea because these satellites move quickly and have different orbits, making their locations seem random.}

\textcolor{black}{For example, work in \cite{9079921}, the user is in an arbitrary location on Earth. They consider a network of a given number of satellites whose locations are modeled as a BPP on a sphere at a fixed altitude.
To suppress the performance mismatch between a random network and practical constellations, which mostly stems from the uneven distribution of satellites along different latitudes, They define and calculate a new parameter, the effective number of satellites, for every user latitude.
They show that, with the above compensation, the generic performance of large deterministic constellations can be very accurately analyzed with theoretical expressions that are based on stochastic constellation geometry.}

\textcolor{black}{The authors of \cite{9218989} investigate the probability of user coverage in LEO satellite communication systems, focusing on scenarios where gateways serve as relays connecting users with LEO satellites.
However, the objective in \cite{9177073}, is to create a model for how communication works in LEO satellite systems. There are two types of communication links: one between gateways and satellites, and another inter-satellite link between LEO satellites.
For the first type, they want to figure out the distribution of the distance between the nearest satellite and a specific point on Earth. For the second type, they want to know the distribution of the distance between one satellite and its closest neighbor.
To do this, they use the mathematical framework of BPP to find the probabilities for these distances.
However, in the previous studies, the precise outage probability and data throughput of LEO satellite communication systems, when considering shadowed-Rician fading, have not been examined using stochastic geometry.}

\textcolor{black}{In the same system model, the authors of \cite{9678973} study three important distances: i) The distance to the nearest satellite; ii) The distance to the satellite that is actively serving the fixed terminal with its main antenna lobe; iii) The distance to the satellite that is actively serving the fixed terminal with its side antenna lobe. 
These distance distributions are crucial for conducting stochastic geometry-based analyses. Unlike previous studies \cite{9079921} and \cite{9218989}, this research takes into account how the satellite antenna's beam pattern affects the overall system performance.}

\subsubsection{Performance Analysis Based on Other Factors}

\textcolor{black}{
In this section, we explore research that evaluates the effectiveness of multi-satellite MIMO systems, taking into account various factors beyond just distance. 
}
The work in \cite{peng2021channel} takes into account the movement of the user with a certain speed. 
The study models the time-varying property of the LMS channel using a four-state Markov transition process, with the LOO distribution used for modeling channel coefficients in each state.
In a downlink scenario, signals are transmitted simultaneously from GEO satellites to the user using an Alamouti code scheme. 
The performances of different transmission schemes, including Dual Satellite Dual Polarization (DSDP), Dual Satellite Single Polarization (DSSP), Single Satellite Dual Polarization (SSDP), and SISO, are compared in this study. The authors find that DSDP outperforms all other tested schemes in terms of BER performance. The results are shown in Section \ref{Comparative Analysis of Existing Research}. 

The work presented in \cite{storek2020multi} focuses on a downlink Multi-User MIMO (MU-MIMO) SatCom scenario with two multibeam GEO satellites operating in the $11.5\,\text{GHz}$ frequency band, and two single-antenna users as receivers. 
The paper categorizes and models various sources that lead to CFO, including frequency converters in the space segment, and Low Noise Blocks (LNBs) of users, each of which has its phase noise. They also take into consideration the CFO contribution due to the Doppler effect for each satellite.
The authors propose a time-varying channel matrix $\tilde{H}(t)=R(t)HT(t)$, where $T(t)$ and $R(t)$ are diagonal matrices, accounting for the CFOs and oscillator phase noise at the space segment and user side, respectively. 
The authors assume no ISLs are available, so they consider signal processing at the gateway station to use satellites in a coordinated way. 
Using the proposed precompensation method, the authors compensate for the frequency offsets remotely from the gateway station.
They show that the resulting phase variation follows a Gaussian distribution with $5^\degree$ of standard deviation, enabling MU-MIMO precoding. 
They show that using MU-MIMO precoding allows the transmission of two independent signals to two non-cooperative single antenna receivers.
They further show that the signal Modulation Error Ratio (MER) is improved compared to SISO, and the channel capacity is increased by a factor of 1.9 when using MU-MIMO instead of SISO.

\textcolor{black}{
The scholarly work by Zhiqiang Li et al. \cite{li2022performance} conducts a theoretical analysis of the characteristics of Multi-Access Interference (MAI) and MAI-plus noise within the context of a CDMA system with downlink multi-satellite system, utilizing the Nakagami-$m$ fading model.}
\textcolor{black}{Each satellite} is responsible for one or more cell communication services that do not overlap.
The authors evaluate the PDF of the received signal and obtain the closed-form expression of MAI, considering different fading models. 
Additionally, they
establish the relationship between the total transmission rate, BER, and the number of users to balance the transmission rate and system reliability. 
The authors show that increasing the number of users can improve the system's total transmission rate, but deteriorates the BER performance when the number of satellites is constant due to the reduction in the transmission power per user.
When the number of users is small, the system's total transmission rate improves linearly with the increase in number of users. 
However, for a larger number of users, the transmission rate is enhanced more slowly or even decreases with an increase in the number of users. 
Therefore, it is inappropriate to continuously add to the number of users to improve the system information rate, as this would consume a lot of energy and have only a slight performance improvement. \textcolor{black}{They conclude that} a balance between the transmission rate and the number of users is necessary to ensure the high energy efficiency of the system.

\textcolor{black}{The authors of \cite{ye2023multi} investigate uplink transmission using a lens antenna array directed towards multiple satellites.}
\textcolor{black}{The use of a lens antenna array has become an appealing choice for ground terminals, allowing them to establish simultaneous connections with multiple satellites and enhance coverage, even when facing energy supply constraints.}
These arrays offer a promising solution for efficient multi-satellite uplink transmission \cite{cao2021multi, garg201728}. Unlike other types of multi-beam antennas, they provide wide-angle scanning and cost-effective, energy-efficient, and low sidelobe beamforming \cite{yang2016band}. In \cite{ye2023multi}, a multi-satellite uplink system using lens antennas is explored. The system includes a ground station and single-antenna satellites, utilizing $N_{RF}$ RF chains on the lens antenna array. The study optimizes analog beam selection, digital precoding, and RF chain count for energy efficiency. \textcolor{black}{They conclude that} interestingly, the relationship between energy efficiency and RF chain count isn't straightforward, as having more chains can enhance spectral efficiency while affecting power consumption. The study also examines how satellite positions impact overall system performance.

\textcolor{black}{There are additional works that could be included in this section. However, they have already been surveyed in other surveys. Therefore, we refer readers to these works and their related surveys for more information. These works include \cite{li2021capacity}, \cite{abdelsadek2022distributed}, \cite{richter2020downlink}, and \cite{roper2022beamspace}, which are surveyed in \cite{heo2023mimo}. Additionally, \cite{liolis2007multi} and \cite{Yamashita2005Broadband} are covered in survey \cite{arapoglou2010mimo}, and \cite{Goto2018LEO} is discussed in survey \cite{al2022survey}.}

\subsection{Collaborative Multi-Satellite Systems}
\label{Collaborative multi-satellite systems}
In this section, we will review some of the latest research that delves into the benefits of satellite collaboration via ISLs.

\textcolor{black}{The authors of \cite{10061620} expand upon their prior research in \cite{abdelsadek2022distributed}, encompassing multiple-antenna satellites and multiple-antenna UTs, and derive the spectral efficiency. In this current paper, they perform a comprehensive comparative analysis by comparing the Distributed Massive MIMO (DM-MIMO) Connectivity (DMMC) scenario with two alternative scenarios: Collocated Massive MIMO Connectivity (CMMC), wherein all satellite antennas are concentrated on a single satellite, and Single Satellite Connectivity (SSC), which involves the participation of only one satellite. Simulation results demonstrate that the DM-MIMO approach surpasses the other two scenarios in terms of spectral efficiency. For a more detailed visual representation of this comparison, please refer to simulation figures presented in Section \ref{Comparative Analysis of Existing Research}.}

In SatCom one crucial issue to address is the estimation of the downlink channel. In \cite{abdelsadek2022distributed, 10061620}, the authors utilize the channel reciprocity assumption, using the estimated uplink channel for downlink transmission. However, in SatCom, the longer transmission delays compared to terrestrial communication can lead to the downlink transmission being out of the coherence interval. As a result, relying on the uplink channel assumption for downlink transmission becomes impractical, leading to a channel estimation imperfection called "delayed CSI". Therefore, it is essential to consider delayed CSI in downlink transmission within SatCom systems.

The work of Omid et al. \cite{omid2023spacemimo} investigates an uplink transmission scenario from a handheld user to a cluster of satellites. Their main focus is on utilizing ISLs to enable cooperative signal detection. They examine two cases: one with sharing complete CSI and the other with sharing partial CSI among the satellites. The study compares these cases in terms of capacity, overhead, and BER. Furthermore, the authors analyze the impact of channel estimation errors in both scenarios and propose robust detection techniques to handle channel uncertainty within a certain range. They evaluate the performance of each case and compare it with conventional SatCom systems, where only one satellite can establish a connection with the user. The results demonstrate that the proposed satellite constellation, consisting of 3168 satellites in orbit, achieves a remarkable capacity of 800 Mbps by leveraging cooperation among 12 satellites and utilizing a $500\,\text{MHz}$ bandwidth. In contrast, conventional SatCom approaches with the same system parameters offer a significantly lower capacity of less than $150\,\text{Mbps}$ for the nearest satellite. They also compare the outcomes of two cases examined in this study, as depicted in Fig. \ref{uplinkcooperation}.

\textcolor{black}{In the context of SatCom networks, obtaining instantaneous CSI is a challenging task.}
In this case, the knowledge of a channel's statistical characteristics can assist in determining channel capacity.
\textcolor{black}{Based on this, the authors of \cite{maaref2007joint} and \cite{ratnarajah2004spatially} first state the channel model and then demonstrate how this distribution can be used to establish a PDF-based approach for characterizing the capacity statistics of MIMO Rician and Rayleigh fading channels.}
If we consider $n$ independent complex Gaussian random vectors $\boldsymbol{h}_i$ of dimension $m$, and $\boldsymbol{H}=[\boldsymbol{h}_1,...,\boldsymbol{h}_n]$, with each column having a distribution of $\boldsymbol{h}_i\sim\mathcal{C}\mathcal{N} (\boldsymbol{m}_i , \boldsymbol{\Sigma})$, then, $\boldsymbol{S} = \boldsymbol{H}\boldsymbol{H}^H$ follows a complex non-central Wishart distribution with $n$ degrees of freedom. It can be expressed as $\mathcal{C}\mathcal{W}_m (n, \boldsymbol{\Sigma}, \boldsymbol{M}\boldsymbol{M}^H)$, where $\boldsymbol{M} = [\boldsymbol{m}_1, \boldsymbol{m}_2, ..., \boldsymbol{m}_n]$. The eigenvalue distributions of non-central Wishart matrices are applicable for capacity analysis of MIMO Rician and Rayleigh fading channels.
Assuming that only the receiver has perfect knowledge of the channel realizations, the authors of \cite{maaref2007joint} and \cite{ratnarajah2004spatially} mathematically investigate the distribution of the eigenvalues of complex non-central Wishart matrices.
These distributions are derived from independent non-necessarily identically distributed non-zero mean Gaussian vectors with a scaled-identity covariance matrix, for characterizing the capacity statistics of MIMO Rician and Rayleigh fading channels.
They also express the joint and marginal eigenvalue distributions of $\boldsymbol{S}$ for three cases in \cite{maaref2007joint} as described in Table \ref{tab5}.
\begin{table}
\centering
\caption{DIFFERENT CASES CONSIDERED IN \cite{maaref2007joint}}
\label{tab5}
\begin{tabular}{p{1cm}|p{6.5cm}}
\hline
\rowcolor{morelightGreen}
{Case 1} & {non-central independent but not necessarily identically distributed (non-i.i.d.) Case - $\boldsymbol{M}\boldsymbol{M}^H$ is full rank}\\
\hline
\rowcolor{lightGreen}
{Case 2} & {non-central i.i.d. Case $\boldsymbol{M}\boldsymbol{M}^H$ is of rank 1}\\
\hline
\rowcolor{morelightGreen}
{Case 3} & {central i.i.d. case - $\boldsymbol{M}\boldsymbol{M}^H = \boldsymbol{0}_m$}\\
\hline
\end{tabular}
\end{table}

Using the results in the above work, Omid et al. in their other work \cite{omid2023oncapacity}, focus on scenarios where the satellite cluster lacks real-time CSI. 
Instead, they rely on sharing statistical CSI between satellites and design a joint detection scheme that minimizes the Mean Square Error (MSE). They calculate the ergodic capacity using the properties of the Wishart matrix that are investigated in \cite{maaref2007joint} and provide a closed-form approximation for low SNR scenarios. Numerical results demonstrate the effectiveness of their proposed detection scheme and the proximity of the approximation to the actual ergodic capacity.
These findings further highlight the potential of utilizing cooperative signal detection and statistical CSI in SatCom systems, offering improved capacity compared to conventional approaches.

\textcolor{black}{In \cite{9814655}, the paper addresses a downlink scenario involving multiple LEO satellites and users in a terrestrial cellular network. Given the unavailability of instantaneous CSI, the study utilizes the Deterministic Equivalence (DE) approach, tailored for large-scale SatCom systems, to derive asymptotically optimal precoding vectors, relying solely on statistical CSI.
These precoding vectors, designed for satellite cooperation, aim to minimize interference within and between cells. The research findings indicate that the DE approach achieves performance levels similar to the Weighted Minimum Mean Square Error (MMSE) approach, which relies on instantaneous CSI. Furthermore, the DE approach significantly outperforms a non-cooperative multi-satellite system.
}

\subsection{Collaborative Multi-Small Satellite Systems}
When it comes to satellite size, the development of satellite technology shows two clear trends. Firstly, there is a noticeable shift towards larger and heavier single satellites that boast complex structures and advanced functionalities. On the other hand, there is a growing popularity of smaller satellites with diverse structures, offering simpler functions achieved through collaborative efforts. This trend aims to replace the complexity associated with a single large satellite. The development of large satellites is limited due to the intricate technology involved, long development cycles, and high costs, as emphasized by Euroconsult \cite{euroconsult}. Small satellites typically refer to satellites weighing less than $500\,\text{kg}$. They can be further categorized into different classes based on their weight. Minisatellites range from 100 to $500\,\text{kg}$, microsatellites range from 10 to $100\,\text{kg}$, nanosatellites range from 1 to $10\,\text{kg}$, picosatellites range from 0.1 to $1\,\text{kg}$, and femtosatellites weigh less than $100\,\text{g}$.

In the case of small satellites, since the whole system completes a space mission through the coordination of a number of small satellites, the design and manufacture of those small satellites can be done using standardized processes and the cost of production becomes lower. Due to the small size and light weight of the small satellites, their launch costs will be greatly reduced. In addition, when a small satellite in the system fails, it can be replaced with a low cost in a short period of time and then the maintenance cost of the entire system is reduced. In short, the adoption of networked small satellites to replace an original large satellite can reduce the total cost of space missions significantly. Moreover, As the networked satellite formation consists of multiple satellites, the information and resource redundancy considered in the system design can enhance the robustness and fault tolerance of the system. At the same time, the parallel and distributed nature of a networked formation system can improve the efficiency of the whole system. Moreover, if the space environment and tasks are more complex or a small satellite in the system is damaged, only a few links related to it will be affected and the whole system will not collapse \cite{liu2018survey}. 

The advantages of a distributed small satellite approach for direct connectivity are highlighted in Avellan's AST SpaceMobile patent \cite{avellan2018system}. This patent introduces the concept of a High Throughput Fractionated Satellite (HTFS) system, where the functions of a conventional monolithic spacecraft are divided among multiple small or very small satellites, in addition to a central command and relay satellite. In this system, the coordinated array of small satellites functions together as a large aperture in space. This distributed satellite aperture effectively enhances the equivalent antenna aperture of the HTFS. As a result, the size demands for RF components, batteries, solar panels, and power handling components are significantly reduced or even eliminated, similar to the waveguide systems employed in monolithic satellites. This size reduction directly leads to a decrease in weight and cost for the satellite system. Additionally, the distributed approach facilitates high throughput capabilities by enabling the spatial reuse of the spectrum.

Authors in \cite{bacci2023formation}, aim to determine an optimal geometric configuration for a satellite formation. This configuration involves finding the appropriate arrangement of individual (sub)antennas carried by each spacecraft, resulting in a desirable radiation pattern for the Formation of Array (FoA). The objective is to identify a geometric layout and antenna distribution that yield a radiation pattern that meets the desired criteria. To implement this radiating architecture, the satellites must be positioned near each other, forming a coherent formation of flying objects. This formation enables effective coordination and control of the satellites as a unified entity. The objective of their research is to demonstrate the utilization of this emerging technology to enhance network throughput in a multibeam S-band mobile communication system, whether through LEO or GEO satellites. The aim is to provide evidence that delivering 5G-like communication services to handheld users is indeed feasible.

They consider multiple satellites equipped with regular planar arrays organized in a square-shaped formation.  Their simulation results explore various array spacing configurations and illustrate that increasing the spacing leads to a narrower main beam. However, this also results in the emergence of high grating lobes which has been studied in the literature \cite{zhang2021research, tenuti2017innovative, piattella2017vertigo}.
Additionally, they analyze the effect of the number of satellites on the performance of the FoA. They find that increasing the number of satellites enhances the FoA pattern, as it boosts the array gain and narrows the central beam without introducing additional grating lobes. However, it is important to note that this increase in performance comes at the expense of increased system complexity and cost. Based on their research, the authors conclude that the FoA solution is not as advantageous for LEO systems compared to GEO systems. This is primarily due to the significantly smaller antenna aperture required for LEO systems, which does not provide a clear benefit over the larger deployable phased array solution typically used in GEO systems.

Regarding this matter, the authors of \cite{tuzi2023satellite} express their belief in the significant potential of satellite swarms specifically in the LEO. Their research focuses on the application of satellite swarms for direct-to-cell connectivity use cases. Satellite swarms are a type of configuration within DSSs. They consist of several small and lightweight satellites, such as CubeSats, equipped with commercial low-gain patch antennas. Swarms offer a flexible and cost-effective solution for satellite systems aimed at direct connectivity, reducing both design and launch expenses \cite{saeed2020cubesat, hadaegh2014development}. The primary goal of this research is to design antenna arrays based on satellite swarms. The study extensively analyses the influence of crucial parameters, including the number of satellites in the swarm, their relative distance, and the array geometry. The researchers demonstrate that increasing the spacing between array elements can result in the occurrence of grating lobes. However, they also propose that optimized array geometries can effectively mitigate this problem.

Satellite swarms offer a significant advantage over traditional large satellites with antenna arrays in terms of higher spatial separation among transmit antennas. This characteristic holds the potential for a substantial boost in spectral efficiency \cite{budianu2015swarm}. Furthermore, the utilization of multiple low-cost satellites in swarms provides enhanced flexibility and scalability compared to relying on a single satellite connected to a station on the ground\cite{verhoeven2011origin}. However, achieving the benefits mentioned earlier would necessitate the availability of instantaneous perfect CSI and robust cooperation among the satellites.

The cooperative behavior of satellite formation systems addresses the limitation of traditional satellite LoS channels, where only power gain is obtained due to the strong correlation between antennas. Using a satellite in a formation can be considered as a massive virtual antenna array. As a result, an additional diversity gain can be achieved. This breakthrough improves overall performance and enhances the characteristics of the SatCom channels \cite{zhang2011generalized}.

In this regard, the authors of \cite{deng2021ultra}, investigate a downlink multi-terminal satellite system operating in an ultra-dense LEO. This system involves multiple satellites flying in a coordinated formation to provide data transmission services to several ground stations. The authors focus on the connectivity between stations and satellites within their respective visibility regions at a specific moment. Their objective is to examine the impact of satellite distribution and formation size on the capacity performance, based on theoretical capacity analysis. The study reveals that the capacity of the system exhibits an initial linear growth as the satellite formation size increases. These results are shown in section \ref{Comparative Analysis of Existing Research}, Fig. \ref{satelliteformationsize}. However, after reaching a certain size threshold, the rate of capacity increase becomes slower. Additionally, the authors demonstrate that as the altitude of the satellite formation increases, the channel capacity initially rises with the number of accessible ground stations. However, it eventually starts to decline due to increased path loss.

To assess uplink transmission in a satellite swarm, authors of \cite{roper2023robust} introduce a new, straightforward precoding technique. This technique enables signal transmission from a VSAT equipped with $N_T$ antennas to a satellite swarm arranged in a triangular configuration, based on imperfect knowledge of the satellite’s positions.
The study focuses on a carrier frequency of $30\,\text{GHz}$ and satellite altitude of $600\,\text{km}$, using the pure LoS channel model.
Furthermore, they use numerical analysis to evaluate how system performance changes based on the distance between satellites. Their simulations also demonstrate that their proposed precoding approach achieves optimal capacity when satellites are positioned far apart and perfect satellite position information is available to the VSAT.

\section{Comparative Analysis of Existing Research}
 \label{Comparative Analysis of Existing Research}
In this section, our objective is to present the simulation results from selected papers, which provide valuable insights into the benefits of employing multi-satellite MIMO systems.
\textcolor{black}{
For details on the parameters used to generate each figure, please refer to Table \ref{tab7}, which provides a comprehensive list of parameters employed in simulating figures found in the related references.
} 

\begin{table*}
\centering
\caption{PARAMETERS IN SIMULATIONS}
\label{tab7}
\begin{tabular}
{p{4.7cm}|p{2cm}|p{2cm}|p{2cm}}
\hline
\rowcolor{lightGreen}
 & Fig. \ref{multiantennamultisatellite} \cite{10061620} & Fig. \ref{satelliteformationsize} \cite{deng2021ultra} & Fig. \ref{uplinkcooperation} \cite{omid2023spacemimo}\\
\hline
\rowcolor{morelightGreen}
\textbf{Noise Power Spectral Density $(N_0)$} & $-174\,\text{dBm/Hz}$& $-203\,\text{dBm/Hz}$ & -\\
\hline
\rowcolor{lightGreen}
\textbf{{Carrier  Frequency  $(f_c)$}} &$2\,\text{GHz}$ & $20\,\text{GHz}$ & $6\,\text{GHz}$\\
\hline
\rowcolor{morelightGreen}
\textbf{Satellite Altitude $(h_{sat})$} & $500\,\text{Km}$& $900\,\text{Km}$& $540\,\text{Km}$, $550\,\text{Km}$\\
\hline
\rowcolor{lightGreen}
\textbf{Wavelength $(\lambda)$} & $15\,\text{cm}$ & $1.5\,\text{cm}$& $5\,\text{cm}$\\
\hline
\rowcolor{morelightGreen}
\textbf{Bandwidth $(BW)$} & - & - & $500\,\text{MHz}$\\
\hline
\rowcolor{lightGreen}
\textbf{Satellite Antenna Gain $(G_{sat})$} & $30\,\text{dB}$& - & $35\,\text{dB}$\\
\hline
\rowcolor{morelightGreen}
\textbf{User Antenna Gain $(G_{user})$} &$0\,\text{dB}$ & - & $5\,\text{dB}$\\
\hline
\rowcolor{lightGreen}
\textbf{Transmitted Power $(P_T)$} & $15\,\text{dBW}$& - & $-2\,\text{dBW}$\\
\hline
\end{tabular}
\end{table*}

\begin{figure}[t]
    \centering
\includegraphics[trim=0cm 0cm 0cm 1.2cm, clip, scale=0.33]{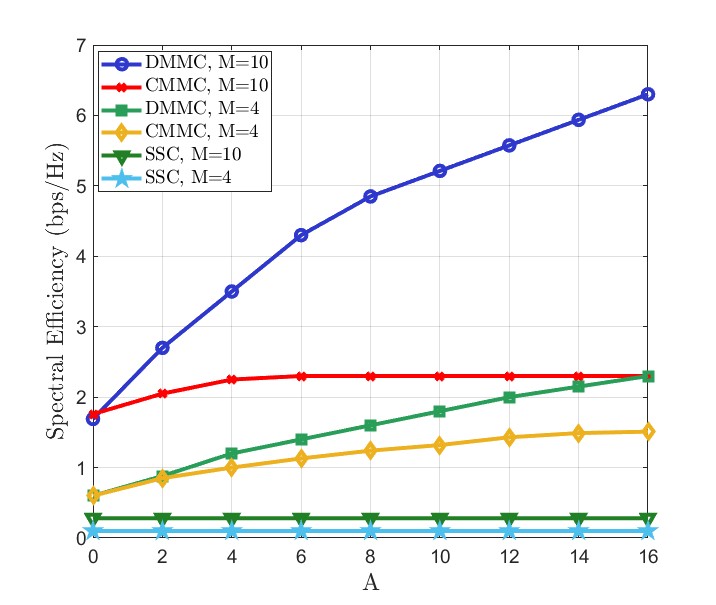}
\caption{Spectral efficiency versus the number of antennas per satellite ($A$) using the three connectivity scenarios at different values of $M$ and the number of antenna at each UT ($N = 2$). \cite{10061620}.}
\label{multiantennamultisatellite}
\end{figure}

\textcolor{black}{
Through the Fig. \ref{multiantennamultisatellite}, readers can gain a clearer understanding of the benefits associated with employing multi-satellite MIMO systems, as opposed to single-satellite single-antenna setups or single-satellite multi-antenna configurations.
}

\textcolor{black}{Fig. \ref{multiantennamultisatellite}, compares spectral efficiency versus a number of antenna per satellite, across three connectivity scenarios described below.}
\begin{itemize}
  \item  \textcolor{black}{ \textbf{DMMC:} A cluster of satellites, each with $A$ antennas, cooperatively transmits to UTs in a DM-MIMO system. The serving satellites transmit the same symbol to the UTs.
    \item \textbf{CMMC:} In this scenario, a single satellite replaces the satellite cluster, with $A^{'} = MA$ antennas and a maximum power of $MP_{max}$, similar to DMMC but with collocated antennas on one satellite.
    \item \textbf{SSC:} This baseline scenario uses SISO mode, where each UT with one antenna connects to a single satellite with one antenna.}
\end{itemize}

\textcolor{black}{ Notably, in the CMMC scenario, adding more antennas does not yield significant gains; The data rate improvement saturates rapidly with a small number of antennas. In contrast, the DMMC scenario benefits from additional antennas, leading to enhanced data rates. Spectral efficiency in the DMMC architecture rises with more satellites ($M$), significantly outperforming the CMMC and SSC scenarios. 
This enhancement is attributed to the deployment of distributed antennas, which introduce macro-diversity and favorable channel conditions. The channel matrix's low spatial dimensionality arises from nearly identical channel paths between the UT's antennas and those at the satellite, owing to substantial distance and a dominant LoS component.}

\begin{figure}[t]
    \centering
\includegraphics[trim=0cm 0cm 0cm 0.12cm, clip, scale=0.51]{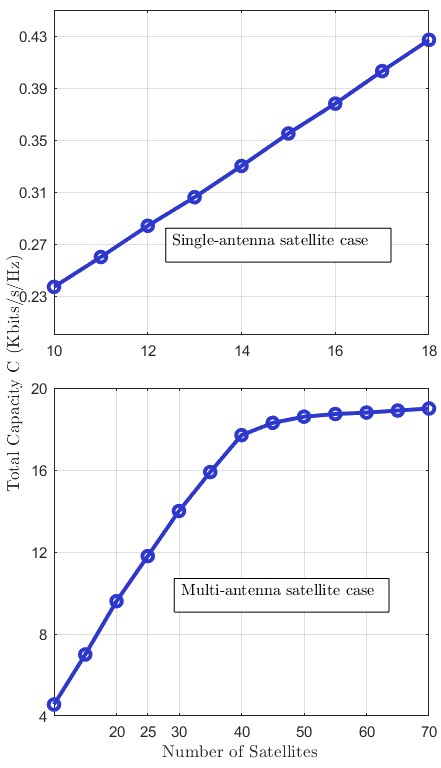}
\caption{Total capacity in two cases versus the satellite formation size \cite{deng2021ultra}.}
\label{satelliteformationsize}
\end{figure}

Fig. \ref{satelliteformationsize} displays the total channel capacity for single-antenna and multi-antenna satellite scenarios relative to satellite formation size, based on work in \cite{deng2021ultra}. Simulation conditions involve a carrier frequency of $20\,\text{GHz}$, $400$ users ($N$) with $2$ antennas each ($K_r$), and $M$ satellites with $20$ antennas each ($K_t$). \textcolor{black}{Furthermore, they introduce a parameter denoted as $\rho$, defined as $\rho = P_TG_TG_R$, which equals $22.4\,\text{dB}$. }
Results reveal a linear capacity increase concerning $M$ when $MK_t<NK_r$, confirming the theoretical analysis in this work. Conversely, when $MK_t$ exceeds $NK_r$, multi-antenna satellite capacity exhibits logarithmic growth with $M$. This behavior occurs because the rank of the channel matrix is determined by the greater number of antennas, whether it is the number of transmit or receive antennas.
Consequently, this observation implies that achieving satisfactory total channel capacity can be attained with a moderately sized satellite formation. Such an approach ensures optimal results without incurring unnecessary costs.



\begin{figure}[t]
    \centering
\includegraphics[trim=0cm 0cm 1.5cm 1.02cm, clip, scale=0.33]{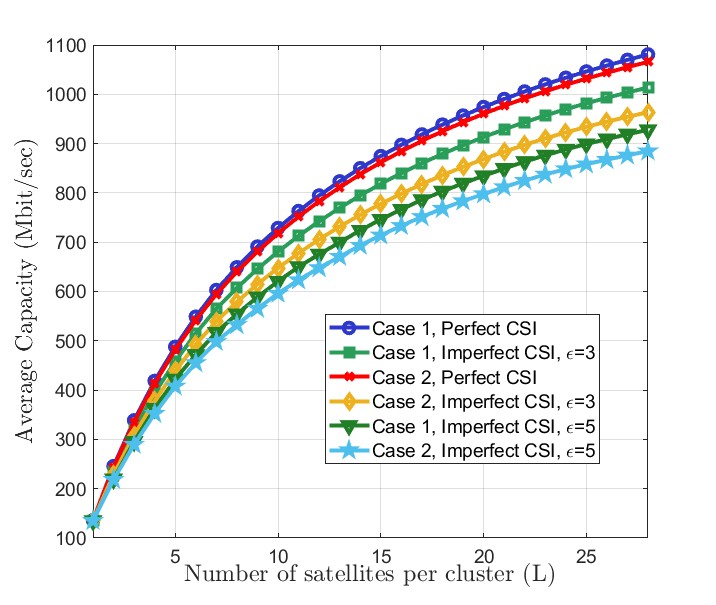}
\caption{Average capacity versus number of satellites in a cluster \cite{omid2023spacemimo}.}
\label{uplinkcooperation}
\end{figure}

\textcolor{black}{
The previously mentioned figures showcased the performance of downlink scenarios in multi-satellite multi-system setups. To display the performance of an uplink scenario in such systems, we refer to the simulation results provided in \cite{omid2023spacemimo}, as follows.}
Fig. \ref{uplinkcooperation} provides valuable insights into the advantages of collaboration among \textcolor{black}{LEO} satellites in terms of system capacity, based on work in \cite{omid2023spacemimo}. As previously noted in subsection \ref{Collaborative multi-satellite systems}, this work involves the consideration of two separate cases: one involves sharing instantaneous CSI among satellites, and the other involves sharing stochastic CSI.
The results show that Case 1, where instantaneous CSI is shared, yields a higher average capacity.
The scenario involves each satellite having a gain of $35\,\text{dB}$, and each user has a gain of $5\,\text{dB}$. The communication is uplink transmission, and the frequency band used is $6\,\text{GHz}$. Note that, the authors of this work, assumed that ISLs are perfectly known and noise-free, but the satellite-user CSI is imperfect and the estimation errors are modeled by Gaussian distributions with zero means and variances equal to $\epsilon^2$ times the variances of the channels.
Based on this figure, by adding more satellites to the cluster of collaborative satellites, the uplink capacity increases, but it does not increase linearly with the number of satellites ($L$), since the satellites are added to the cluster based on their distances to the user in ascending order.
The authors also show in their work that in addition to the number of collaborative satellites, the density of the constellation also affects the capacity. 
With a fixed value for $L$, a dense constellation can provide higher capacity, since the distances of the satellites in the cluster to the user are small compared to a sparse constellation. 

\section{Future Works and Open Challenges}
\label{FutureWorks}
According to the findings from the previous review conducted in this study, several challenges require attention and resolution in this research domain. These issues are outlined as follows:
\begin{itemize}
    \item \textbf{Link Budget:}
    \textcolor{black}{
    With the advancement of technology, one objective is to address the path loss issue in the link budget. This problem can be resolved using highly directive antennas offering substantial antenna gains. This technological evolution aims to reduce the impact of path loss and enhance the overall system performance, as highlighted in the work by Gultepe et al. \cite{gultepe2020dual}.
    However, for direct communication between satellites and unmodified users, this might not be an optimal solution, as there are power limitations in unmodified users as well. A multi-satellite system can help compensate for these power constraints, thereby meeting the link budget requirements.}

    \item \textbf{Delayed CSI:} One of the challenges in SatComs is channel estimation. Some research papers such as \cite{abdelsadek2022distributed} focus on a Time Division Duplex (TDD) scenario in SatComs and under the assumption of channel reciprocity, they estimate the CSI. Although this is a valid assumption for terrestrial communications, in SatComs it is highly impractical. The large distance between satellites and users causes delays in channel estimation and the estimated channel is no longer reliable to be used for the downlink precoding. This issue which is referred to as the delayed CSI problem is one of the main challenges that needs to be addressed. In \cite{zhang2021deep} this problem is addressed in massive MIMO (mMIMO) satellite systems. The authors use the correlation properties of the channels to provide a proper estimation of CSI in the downlink using machine learning. However, this approach cannot be used in distributed satellite systems where multiple satellites far away from each other collaborate on signal processing since there are no CSI correlation properties. Thus, this remains an open problem which needs further investigation.
     \item \textbf{ISL noise:} When sharing information in the ISLs, the receiver noise can impact the reliability of the transmission, especially in cases where due to limited sources the transmission power is limited. The effect of ISL noise on collaborative multi-satellite networks is essential to be investigated and robust techniques should be presented to cope with this issue. This becomes more severe when the bandwidth of ISLs is wide which is the case for THz ISLs and FSO ISLs. Although the throughput increases with larger bandwidths, the power of additive noise also rises which escalates this problem \textcolor{black}{\cite{9108615}}.
    \item \textbf{Time Synchronization:}
    \textcolor{black}{The cooperative transmission of LEO satellites requires synchronization among them.} 
    Accurate time synchronization is essential for scheduling satellite operations, facilitating data transmission, and managing handover procedures. 
    For example, in satellite constellations with multiple satellites, precise time synchronization is vital for seamless satellite handoffs as users move across different coverage areas \textcolor{black}{\cite{10286242}}.
    \textcolor{black}{This synchronization is challenging due to the movement of the satellites.
    Although there are some researches to address this challenge. For instance, in \cite{4540557}, the authors propose a method to establish time synchronization in distributed satellite systems. Additionally, references \cite{6214424} and \cite{8585093}, have explored both closed-loop and open-loop phase and frequency synchronization techniques. These methods offer effective means to achieve synchronization among the satellites.
    }
    
    \item \textbf{Doppler Shift Compensation:} In SatComs, the relative motion between the satellite and the users causes a Doppler shift in the transmitted signal. Doppler compensation in SatComs is essential for correcting the frequency shift caused by the relative motion between the satellite and the users. By applying Doppler compensation, communication systems can ensure proper frequency alignment and maintain continuous communication during satellite movement across the sky. 
    \textcolor{black}{
    As LEO satellites traverse their orbits, their relative velocities change continuously, resulting in dynamic Doppler shifts that affect signal reception and transmission. In the context of multi-satellite systems, the Doppler shift experienced by LEO satellites varies significantly across different orbits and time intervals. This variability adds a layer of complexity to the task of accurately estimating and compensating for these shifts which are essential to ensure reliable communication performance in such environments \cite{pedrosa2021state}.}
    
    \item \textbf{Interference Mitigation:} Managing interference is crucial for the successful operation of multi-satellite systems, especially in densely populated satellite constellations.
    \textcolor{black}{Different types of interference are MAI, Co-Channel Interference (CCI), and CCI between systems. While Spread Spectrum (SS) techniques offer resilience against many types of interference, they can still be affected by the near-far effect caused by MAI. To mitigate this, strategies like power control, multi-user detection, and careful selection of SS codes are employed \cite{686783}. Furthermore, in multi-beam satellite systems, there can be CCI concerns, particularly in cases where adjacent beams share the same frequency due to the angular side-lobes of beam radiation patterns \cite{7105655}.
    Additionally, there has been a growing deployment of LEO satellites in recent years. However, the available radio spectrum remains limited, necessitating the need for LEO SatComs to achieve high spectral efficiency to address this spectrum scarcity issue. Furthermore, GEO SatComs must coexist within the same spectrum to attain their objectives. Consequently, a high level of CCI between LEO and GEO SatComs is inevitable. Current ITU regulations \cite{9347950}, mandate the consideration of spectrum sharing between GEO and LEO SatComs, with GEO SatComs designated as the Primary User and LEO SatComs as the Secondary User. Therefore, effective interference coordination is crucial to mitigate such interference challenges \cite{10209551}.}
    
    \item \textbf{Handover Management:}
    \textcolor{black}{Handover management poses a significant challenge in LEO SatComs due to the high mobility of satellites and unpredictable user movements. Frequent beam realignments and handovers result in signal delays, increased costs, and poor user service experiences. Addressing this, the handover strategy can be improved by considering both single-user and group handovers. For single-user handover issues, advanced deep reinforcement learning methods can be applied to handover decisions. Group handovers, involving multiple users initiating handovers simultaneously as satellites move rapidly, require strategies to reduce signaling congestion and minimize conflicts. 
    To prevent too many handover requests on satellites, we can group users based on where they are or what they do. Then, we can pick one representative carrying the information of other users in each group. Also, when lots of users switch together, we can use reinforcement learning to make the handover work better in these tricky situations \cite{9970355}.}
    \item \textbf{Constellation Design:}
    \textcolor{black}{The geometric arrangement of satellites in a constellation impacts system performance. Designing constellations with the right geometry to support multi-satellite MIMO operations while meeting specific communication needs is a challenge. Moreover, designing and managing ISLs in multi-satellite constellations is essential for enabling cooperation between satellites. Efficient ISL design, considering factors like link quality, power requirements, and data routing, is an ongoing challenge \cite{muttiah2023satellite}.}
    \item \textbf{Long propagation delay of integrated satellite-terrestrial network:}
    \textcolor{black}{The integration of satellite and terrestrial networks presents a promising architecture for offering global broadband access to diverse users. However, a key obstacle in integrating satellite networks with terrestrial ones lies in the extended propagation delay characteristic of SatCom. While propagation delay in LEO/MEO setups can potentially be minimized to tens of milliseconds, the multihop transmission in MEO/LEO satellite networks inevitably boosts propagation delay due to additional ISLs or user-satellite link connections. Consequently, despite the anticipated increase in connectivity resulting from the integration of satellite and terrestrial networks, communication latency remains a significant challenge. Addressing this latency issue is crucial for enhancing the QoS for users \cite{9610113}.}

\end{itemize}

\section{Conclusions}
\label{Conclusion}
The primary objective of this survey is to investigate direct user-satellite communication. To achieve this, the paper presents a comprehensive analysis of multi-satellite MIMO systems, which serve as a viable approach to enable direct user-satellite communication.
The discussion begins with an overview of SatComs, covering its structure, various links utilized in SatComs, suitable frequency bands, and satellite antenna design. Additionally, it includes a comprehensive review of different ISLs, particularly focusing on hybrid FSO/RF ISLs, with relevant works highlighted in this area.
Furthermore, the survey investigates channel models used in SatComs, covering both the satellite-satellite link and user-satellite link channel models, with a particular emphasis on those employed in OTFS-based satellite systems. 
The survey emphasizes the advantages of employing multi-satellites instead of relying on single-antenna single satellite or multi-antenna single satellite. This exploration is supported by an in-depth examination of various studies within the field of multi-satellite MIMO systems. Additionally, the survey explores research on satellite collaboration to enhance system performance. It also discusses the benefits of multi-small satellites compared to larger satellites.
Moreover, simulation results from state-of-the-art research reinforce the benefits of utilizing multi-satellite MIMO systems. In conclusion, this survey provides valuable insights into the potential of multi-satellite MIMO systems to improve the capabilities of direct communication between satellites and users.

\bibliographystyle{unsrt}
\bibliography{Reference}


\begin{IEEEbiography}[{\includegraphics[width=1in,height=1.25in,clip,keepaspectratio]{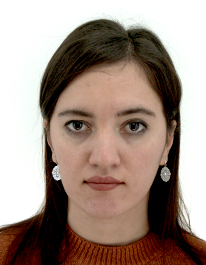}}]{\textbf{Zohre Mashayekh Bakhsh}}
    completed her B.Sc. degree in Electrical Engineering - Telecommunication at Babol Nooshirvani University of Technology, Iran, in 2015. She pursued her master's degree at K.N. Toosi University of Technology, Iran, focusing on the application of beamforming in cognitive radio systems and MIMO relay networks. She is currently working on the “Space MIMO” project as a Ph.D. student at the 6GIC of the University of Surrey. Her research revolves around multi-satellite MIMO systems to enable direct mobile-satellite communications. Her research interests encompass Satellite Communications (SatComs), MIMO technology, and Wireless Communications.
\end{IEEEbiography}

\begin{IEEEbiography}[{\includegraphics[width=1in,height=1.25in,clip,keepaspectratio]{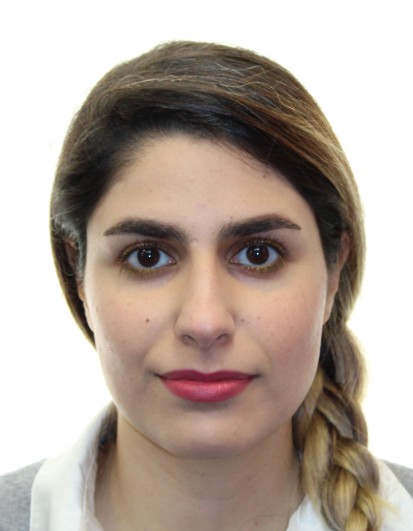}}]{\textbf{Yasaman Omid}}
received a Bachelor’s degree in Electronics Engineering from the University of Isfahan,  and a Master’s degree from K. N. Toosi University of Technology in Telecommunications Engineering with a focus on Massive MIMO systems.
She completed a PhD program in Telecommunications Engineering at Queen Mary University of London with a focus on Intelligent Reflecting Surfaces. She worked as a postdoctoral research fellow on the Space MIMO project at the University of Surrey. Currently, she is a postdoctoral research associate at Loughborough University, working on space-air-ground integrated networks.
\end{IEEEbiography}

\begin{IEEEbiography}[{\includegraphics[width=1in,height=1.25in,clip,keepaspectratio]{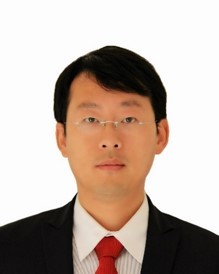}}]{\textbf{Gaojie Chen}}
received the B.Eng. and B.Ec. Degrees in electrical information engineering and international economics and trade from Northwest University, China, in 2006, and the M.Sc. (Hons.) and PhD degrees in electrical and electronic engineering from Loughborough University, Loughborough, U.K., in 2008 and 2012, respectively. After graduation, he took up academic and research positions at DT Mobile, Loughborough University, University of Surrey, University of Oxford, and University of Leicester, U.K. He is currently an Assistant Professor with the Institute for Communication Systems, 5GIC \& 6GIC, University of Surrey, U.K., and a Visiting Research Collaborator with the Information and Network Science Lab, University of Oxford, U.K. His current research interests include information theory, wireless communications, satellite communications, cognitive radio, the Internet of Things, secrecy communications, and random geometric networks. He received the Exemplary Reviewer Awards of the {\scshape IEEE Wireless Communications Letters} in 2018, the {\scshape IEEE Transactions on Communications} in 2019 and the {\scshape IEEE Communications Letters} in 2020 and 2021; and Exemplary Editor Awards of the {\scshape IEEE Communications Letters} and {\scshape IEEE Wireless Communications Letters} in 2021 and 2022, respectively. He served as an Associate Editor for the {\scshape IEEE Journal on Selected Areas in Communications - Machine Learning in Communications} from 2021-2022. He serves as an Associate Editor for the {\scshape IEEE Communications Letters}, {\scshape IEEE Wireless Communications Letters} and {\it Electronics Letters} (IET).
\end{IEEEbiography}

\begin{IEEEbiography}[{\includegraphics[width=1in,height=1.25in,clip,keepaspectratio]{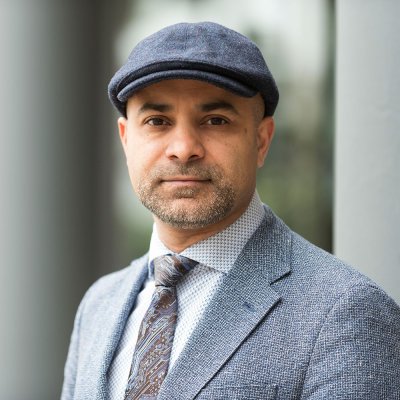}}]{\textbf{Farbod Kayhan}}
    received the B.Sc. degree in mathematics from the S.B. University of Kerman, Iran, in 1999, the M.Sc. degree in modeling and simulation of complex realities from the International Centre for Theoretical Physics, Trieste, Italy, in 2003, and the Ph.D. degree in electrical engineering and telecommunication from Politecnico di Torino, Italy, in 2007. He has spent a one-year Postdoctoral position with the Department of Systems Science of Kyoto University from 2007 to 2008. He was a Research Assistant with Politecnico di Torino from 2009 to 2015 and a Research Associate with the SIGCOM group at, the University of Luxembourg from 2015 to 2020. He was a Lecturer of satellite communication at the 5G and 6G innovation center at the University of Surrey from 2020 to 2023. He is currently a lead satcom 5G R\&D engineer at Telespazio, UK. His main research interests include modern coding theory and combinatorial optimization with particular emphasis on iterative channel and source coding and constellation space optimization, signal processing, and satellite communication systems.
\end{IEEEbiography}

\begin{IEEEbiography}[{\includegraphics[width=1in,height=1.25in,clip,keepaspectratio]{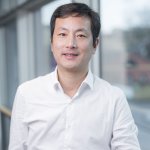}}]{\textbf{Yi Ma}}
    is a professor with the Institute for Communication Systems (ICS), formerly the Centre for Communication Systems Research (CCSR), University of Surrey, Guildford, U.K. He has extensive expertise in the areas of signal processing, machine learning, and information theory, with their applications in telecommunications.
\end{IEEEbiography}


\begin{IEEEbiography}[{\includegraphics[width=1in,height=1.25in,clip,keepaspectratio]{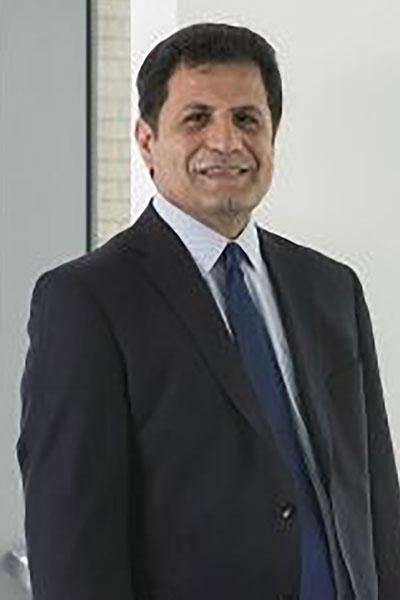}}]{\textbf{Rahim Tafazolli}}
    (Senior Member, IEEE) is a Regius Professor of Electronic Engineering, a Professor of Mobile and Satellite Communications, and the Founder and the Director of 5GIC, 6GIC, and the Institute for Communication System (ICS), University of Surrey, Guildford, U.K. He has more than 30 years of experience in digital communications research and teaching. He has authored or co-authored more than 1000 research publications and is regularly invited to deliver keynote talks and distinguished lectures at international conferences and workshops.
\end{IEEEbiography}

\end{document}